\documentclass[superscriptaddress,twocolumn,showkeys,
amssymb,amsmath,nobibnotes,aps,prd,
showkeys, nofootinbib]{revtex4-1}
\pdfoutput=1
\usepackage{graphicx,bm,color,psfrag,hyperref}
\usepackage{amsfonts}
\usepackage{lipsum}
\usepackage{mathtools}
\usepackage{verbatim}
\usepackage[dvipsnames]{xcolor}

\definecolor{navyblue}{rgb}{0.0, 0.0, 0.5}
\definecolor{royalblue}{rgb}{0.25, 0.41, 0.88}
\definecolor{cadmiumgreen}{rgb}{0.0, 0.42, 0.24}
\definecolor{blue-violet}{rgb}{0.54, 0.17, 0.89}
\definecolor{darkviolet}{rgb}{0.58, 0.0, 0.83}
\definecolor{orange(colorwheel)}{rgb}{1.0, 0.5, 0.0}

\hypersetup{
	colorlinks=true, 
	linkcolor=royalblue, 
	citecolor=magenta}

\usepackage{siunitx} 
\usepackage{scalerel} 
\usepackage{enumitem}

\newcommand{\ie}{\textit{i.e.}\,}

\begin{document}

\title{Cosmological constraints on slow roll inflation: An update}

\author{Matteo Forconi}
\email{forconi.1715343@studenti.uniroma1.it}
\affiliation{Physics Department and INFN, Universit\`a di Roma ``La Sapienza'', Ple Aldo Moro 2, 00185, Rome, Italy} 

\author{William Giar\`e}
\email{william.giare@uniroma1.it}
\affiliation{Physics Department and INFN, Universit\`a di Roma ``La Sapienza'', Ple Aldo Moro 2, 00185, Rome, Italy} 

\author{Eleonora Di Valentino}
\email{eleonora.di-valentino@durham.ac.uk}
\affiliation{Institute for Particle Physics Phenomenology, Department of Physics, Durham University, Durham DH1 3LE, United Kingdom}

\author{Alessandro Melchiorri}
\email{alessandro.melchiorri@roma1.infn.it}
\affiliation{Physics Department and INFN, Universit\`a di Roma ``La Sapienza'', Ple Aldo Moro 2, 00185, Rome, Italy} 

\date{\today}

\preprint{}
\begin{abstract}
In light of the most recent cosmological observations, we provide new updated constraints on the slow roll inflation in different extended scenarios beyond the $\Lambda\rm{CDM}$ cosmological model. Along with the usual six parameters, we simultaneously vary different combinations of additional parameters, including the running of the scalar spectral index $\alpha_s$, its running of running $\beta_s$, the tensor amplitude $r$ and the spatial curvature $\Omega_k$. From the Planck 2018 data, we find no evidence for a scalar running or a running of running, while analyzing the Atacama Cosmology Telescope data combined with WMAP 9-years observations data we find a preference for nonzero $\alpha_s$ and $\beta_s$ at the level of 2.9$\sigma$ and 2.7$\sigma$, respectively. Anyway, this preference is reduced when the tensor amplitude can vary in the model or $\beta_s$ is fixed to zero. The upper bound on $r$ is only slightly affected by the additional parameters while the differences in the datasets can remarkably change the compatibility among the different inflationary models, sometimes leading to discordant conclusions. 
\end{abstract}

\maketitle


\section{Introduction} \label{sec:intro} 
In the very early Universe a phase of almost de Sitter expansion known as cosmological inflation is supposed to set the initial condition for hot big bang theory evolution, driving the Universe toward homogeneity and flatness~\cite{Guth:1980zm}.

The simplest dynamical model of inflation involves a  scalar field $\phi$, which from now on we call the inflaton, minimally coupled to gravity. The action of the theory reads 
\begin{equation}
S=\int d^4x\,\sqrt{-g}\left[\frac{M_p^2}{2}\,R+\frac{1}{2}\,g^{\mu\nu}\partial_{\mu}\phi\partial_{\nu}\phi-V(\phi)\right],
\label{minimal coupled action}
\end{equation}
with $M_p=1/\sqrt{8\pi G}=2.4\times 10^{18}\,\rm{GeV}$ the reduced Planck mass in the natural units ($c=\hslash=1$). Notice that this theory is said to be “minimal coupled to gravity” because there is not a direct coupling between the inflaton field and the metric tensor in the action. The equation of motion of the  field can be obtained minimizing the action with respect to the field $\delta S / \delta\phi=0$. Restricting our attention to homogeneous scalar fields and adopting a flat Friedmann-Lema\^itre-Robertson-Walker (FLRW) metric one obtains
\begin{equation}
\ddot{\phi}+3H\dot{\phi}+V'(\phi)=0.
\label{inflaton equation of motion}
\end{equation}
On the other hand, minimizing the action with respect to the metric, $\delta S/\delta g^{\mu\nu}=0$, one can find the relation for the energy-density $\rho_{\phi}$ and pressure $P_{\phi}$ in an inflaton-dominated Universe, namely:
\begin{equation}
\omega_{\phi}\doteq \frac{P_{\phi}}{\rho_{\phi}}=\frac{\frac{1}{2}\dot{\phi}^2-V(\phi)}{\frac{1}{2}\dot{\phi}^2+V(\phi)}.
\label{inflaton equation of state}
\end{equation} 
Therefore a dynamical scalar field can induce an epoch of expansion provided that the inflationary potential $V(\phi)$ is sufficiently flat to allow a phase of slow-roll evolution \cite{Linde:1981mu,Baumann:2009ds,Martin:2013tda,Kinney:2009vz}. Indeed if $V\gg\dot{\phi}^2$ a phase of repulsive gravity $\omega_{\phi}\approx-1<-1/3$ is obtained and the Universe starts inflating.
It is also easy to see that the background evolution is completely fixed by the inflaton dynamics on the potential:
\begin{equation}
3 M^2_p\, H^2=\rho_{\phi}=\left(\frac{1}{2}\dot{\phi}^2+V(\phi)\right),
\label{inflato freedman 1}
\end{equation}
\begin{equation}
M^2_p\,\left(\frac{\ddot a}{a}\right)=-\frac{1}{6}\left(\rho_{\phi}+3P_{\phi}\right)=-\frac{1}{3}\left(\dot{\phi}^2-V(\phi)\right).
\label{inflaton freedman 2}
\end{equation}
Notice also that, because of Eqs.~\eqref{inflato freedman 1} and \eqref{inflaton freedman 2}, the slow-roll condition is equivalent to require that $V'(\phi)\approx-3\,H\dot{\phi}$ or  $|\dot H|/ H^2\ll 1$. Therefore it is useful to introduce the following potential slow-roll parameters
\begin{subequations}
	\begin{equation}
	\epsilon _ { V } \doteq M _ { \rm { pl } } ^ { 2 } \frac{1}{2} \left(\frac { V _ { \phi } ^ { 2 } } { V ^ { 2 } }\right),
	\end{equation}
	\begin{equation}
	\eta _ { V } \doteq M _ { \rm { pl } } ^ { 2 } \left(\frac { V _ { \phi \phi } } { V }\right),
	\end{equation}
	\begin{equation}
	\xi^2 _ { V } \doteq  M _ { \rm { pl } } ^ { 4 } \left(\frac { V _ { \phi } V _ { \phi \phi \phi } } { V ^ { 2 } }\right),
	\end{equation}
	\begin{equation}
	\varpi _ { V } ^ { 3 } \doteq M _ { \rm { pl } } ^ { 6 }\left(\frac { V _ { \phi } ^ { 2 } V _ { \phi \phi \phi \phi } } { V ^ { 3 } }\right)
	\end{equation}
\end{subequations}
where $V_{\phi\dots\phi}\doteq V^{\prime\dots\prime}$ indicates the derivatives of the potential with respect to the filed. Notice that the potential parameters will be largely used in the subsequent discussion together with the parameters $\{\epsilon_i\}$ defined as 
\begin{equation}
\epsilon_1\doteq-\frac{\dot H}{H^2}\simeq \epsilon_V, \quad \epsilon_{i>1}\doteq \frac{d \log\epsilon_{i-1}}{d\log k}.
\end{equation}
that are instead clearly related to the background dynamics. During the slow-roll phase all these parameters are expected to be small with the limit $1\gg |\epsilon_{V}|\simeq|\epsilon_1|\to0$ corresponding to an exactly de Sitter expansion.

Furthermore, the quantum fluctuations of the field around its classical trajectory, becoming classical on large scales, can induce energy-density fluctuations, sourcing both rotational invariant scalar modes and, if the energy scale of inflation is sufficiently high, a satiable background of primordial gravitational waves (PGWs) \cite{Linde:1981mu,Vilenkin:1983xq,Lyth:2009zz,Mukhanov:2005sc,Starobinsky:1980te,Weinberg:2008zzc,Martin:2013tda,Riotto:2002yw,Caprini_2018,Cabass:2015jwe}.  Being scalar and tensor perturbations decoupled at the linearized level, they can be treated separately.  After the end of inflation, scalar perturbations reenter the observable Universe, setting the seeds for the structure formation and providing a quite natural explanation for the observed anisotropies in the cosmic microwave background (CMB).  On the other hand, PGWs may imprints the CMB photon polarization, leading to a very distinctive signature in the B-modes spectrum on large angular scales~\cite{Guth:1980zm,Starobinsky:1980te,Linde:1981mu,Vilenkin:1983xq,Mukhanov:2005sc,Dodelson:2003ft,Weinberg:2008zzc,Martin:2013tda,Baumann:2009ds,Clarke:2020bil}.  

In the framework of single-field inflation with Einstein gravity, primordial scalar and tensor perturbations are expected to be (nearly) Gaussian and hence they can be described in terms of their two-point correlation functions and their primordial spectra\footnote{We recall that for a generic Gaussian random field $\psi_k$, the spectrum is defined in terms of its two-point correlation function as $\langle \psi_k\,\psi_{k\prime}\rangle\doteq (2\pi)^3\delta^{3}_{k+k\prime} \,P_{\psi}(k)$ and that the other higher-order correlation functions are expected to vanish. Notice also that in the text we work in terms of the \textit{dimensionless} spectra defined as $\mathcal P_{\psi}(k)\doteq (k^3/2\pi^2)\,P_{\psi}(k)$.}.  It is well known that the spectrum of the quantum fluctuations of a (massless) scalar field in a de Sitter background is flat. Therefore, since both scalar and tensor perturbations are sourced by the fluctuations of the inflaton field in an almost de Sitter background, we expect nearly but not exactly flat primordial spectra. As a matter of fact, in the framework of the single field slow-roll approximation~\cite{Lyth:2009zz,Mukhanov:2005sc,Dodelson:2003ft,Weinberg:2008zzc}, from the theory of the quantum inflationary fluctuations~\cite{Guth:1985ya,Starobinsky:1992ts, Starobinsky:1979ty,Starobinsky:1983zz,Mukhanov:1981xt,Mukhanov:1990me,Mukhanov:2013tua,Mukhanov:1982nu,Bardeen:1980kt,Bardeen:1983qw,Adams:1992bn,Bartolo:2001rt,Choe:2004zg,Gordon:2000hv,Jackson:2013vka,Martin:2002vn,Adams:2001vc}, one can calculate the primordial spectra predicted by inflation, obtaining
\begin{equation}
\mathcal{P}_ {s}= \left(\frac{1}{8\pi^2 M_{\rm pl}^2}\right)\left(\frac{H^2}{\epsilon_V}\right)=\left( \frac { 1 } { 12 \pi ^ { 2 } M _ { \rm pl } ^ { 6 } } \right) \left( \frac { V ^ { 3 } } { V_{\phi} ^ { 2 } } \right),
\label{scalar_spectrum}
\end{equation}
\begin{equation}
\mathcal{P}_{T} = \left(\frac{2}{\pi^2 M_{\rm pl}^2}\right)H^2=\left( \frac { 2 } { 3 \pi ^ { 2 } M _ { \rm pl } ^ { 4 } } \right) V.
\label{tensor_spectrum}
\end{equation}
These relations are evaluated when a wave number $k$ exits the causal horizon during inflation, namely $k=a H$. Since by definition the expansion rate is almost constant during the slow-roll evolution ($H^2\gg |\dot{H}|$), the perturbations produced by generic single-field models are typically well approximated by the following power-law form of the adiabatic scalar and tensor components
\begin{equation}
\log \mathcal{P}_{\rm s}(k)=\log A_{\mathrm{s}}+\left(n_{\mathrm{s}}-1\right) \log \left(k / k_{*}\right) + \dots 
\label{PLS}
\end{equation}
\begin{equation}
\log \mathcal{P}_{\rm T}(k)=\log \left(r\,A_{\mathrm{s}}\right)+\left(n_{\mathrm{T}}\right) \log \left(k / k_{*}\right) + \dots
\label{PLT}
\end{equation}
where $k_*$ denotes an arbitrary scale known as \textit{pivot scale}, $A_{\rm s}\doteq \mathcal{P}_{s}(k_*)$ and $A_{\rm T}\doteq \mathcal{P}_{\rm T}(k_*)$ are the scalar and tensor amplitudes computed at the pivot scale and $r\doteq A_{\rm T} / A_{\rm s}$ is the so called tensor-to-scalar ratio. We have also defined the scalar and tensor spectral index (or tilt) respectively as
\begin{equation}
n_{\rm s}-1\doteq \left[\frac{d\log \mathcal{P}_{s}(k)}{d\log k}\right]_{k=k_*},
\end{equation}
\begin{equation}
n_{\rm T}\doteq \left[\frac{d\log \mathcal{P}_{\rm T}(k)}{d\log k}\right]_{k=k_*}.
\end{equation}
The scalar and tensor tilts quantify the departure from the scale-invariant case and in this simplest scenario we expect slightly tilted spectra because of the field evolution which breaks the de Sitter isometries, providing also a well-defined clock to measure the time to the end of inflation. Notice anyway that inflation does not predict neither the precise values of the amplitudes nor those of the tilts, but they depend on the details of the inflationary dynamics which is clearly related to the precise shape of the potential. Indeed, using Eqs.\eqref{scalar_spectrum} - \eqref{tensor_spectrum}, one can write the scalar and tensor tilt in terms of the slow-roll parameters as
\begin{equation}
n_{s}-1=2\eta_V-6\epsilon_V=-2\epsilon_1-\epsilon_2,
\label{Spectral}
\end{equation}
\begin{equation}
n_T=-2\epsilon_V=-r/8
\label{Tensorial}
\end{equation}
from which we see that constraints on the spectral parameters can be translated into constraints on the inflationary potential (or the background dynamics) and vice-versa. Notice also that, within the power-law parametrization, both $n_{\rm s}$ and $n_{\rm T}$ are usually assumed to be scale invariant and the higher order terms in Eqs. \eqref{PLS} and \eqref{PLT} are typically ignored. This is clearly an approximation and a further parametrization that includes also higher-order corrections could be considered \cite{Kuroyanagi:2011iw,Zarei:2014bta,Giare:2020vhn}. 
Moreover, it is worth noting that today only the scalar amplitude and tilt are measured with good precision~\cite{Akrami:2018odb} while a detection of primordial tensor modes is still missing and a combined analysis of the Planck measurements of the CMB polarization and anisotropies \cite{Akrami:2018odb} and the BICEP2/Keck array likelihood for B-modes polarization~\cite{Ade:2018gkx} constrains the amplitude of primordial gravitational waves on scales comparable to the current Hubble length ($a_0\,H_0= 2.248\times 10^{-4}\,\rm{Mpc}^{-1}$), placing only an upper bound on the tensor to scalar ratio at a pivot scale $k_* = 0.002\,\rm{Mpc}^{-1}$ of $r_{0.002} < 0.056$ at $95\%$ Confidence Level (CL hereafter).

Therefore, even though current observations show a general agreement with the standard slow-roll predictions and many inflationary models proposed in literature can be ruled out, it should be noted that the missing evidence for tensor modes and, in general, the present day accuracy in data places only generic constraints on inflation that in many cases are obtained within the specific assumptions of the standard $\Lambda\rm{CDM}$ cosmological model (\textit{e.g.} an exactly flat background geometry, a vanishing scale dependence of the scalar -- and tensor -- tilt or even a negligible tensor amplitude).

In this paper we review and discuss constraints on the slow-roll paradigm considering different extensions of the $\Lambda\rm{CDM}$ model. Because of the inflationary perspective of this work, we are basically interested in exploring modifications in the primordial sector. In light of the most recent cosmological observations, we both update the existing bounds on the (higher-order) inflationary parameters and derive new constraints on the inflationary dynamics in nonstandard scenarios (including the case of a curved cosmological spacetime). By using the slow-roll approximation we also relate the primordial perturbations to the dynamics of the Hubble parameter during inflation or to the inflaton potential and its derivatives, constraining these quantities and interpreting the results in terms of the physics of the inflationary epoch.

The paper is organized as follows. In Sec.~\ref{sec.Methods} we describe our analysis method; in Sec.~\ref{sec.Results} we discuss our results; finally in Sec.\ref{sec.Conclusion} we present our conclusions.

\section{Methodology} \label{sec.Methods} 

In this section we outline the methodology used in our analysis. We start pointing out the cosmological model, highlighting the modifications to the primordial sector and discussing the additional parameters that we introduce. Then, we review our data-analysis techniques and the cosmological datasets used to derive our results.

\subsection{Cosmological model}
We analyze the slow-roll paradigm of inflation considering different extensions of the standard $\Lambda$CDM model. We consider the standard $\Lambda$CDM cosmological model described by the usual six-parameters, i.e., the baryon $\omega_{\rm b}\doteq \Omega_{\rm b}h^2$ and cold dark matter $\omega_{\rm c}\doteq\Omega_{\rm c}h^2$ energy densities, the angular size of the horizon at the last scattering surface $\theta_{\rm{MC}}$, the optical depth $\tau$, the amplitude of primordial scalar perturbation $\log(10^{10}A_{\rm S})$ and the scalar spectral index $n_s$. 

Together with the standard $\Lambda$CDM parameters, we consider different combinations of additional parameters that involve modifications either in the primordial sector or in the spacetime geometry.
In particular we generalize Eqs.\eqref{PLS} to the following expansion
\begin{align}
\nonumber \log\mathcal{P}_{\rm s}(k)=&\ln A_{\mathrm{s}}+\left(n_{\mathrm{s}}-1\right) \log\left(k / k_{*}\right) + \frac{\alpha_s}{2} \log^2 \left(k / k_{*}\right)\\
& +\frac{\beta_s}{6} \log^3 \left(k / k_{*}\right)
\label{PhenomenologicalScalar}
\end{align}
introducing a weak scale-dependence in the primordial spectrum modeled by the running of the scalar tilt $\alpha_s$ or also its running of running $\beta_s$ defined as 
\begin{equation}
\alpha_{\rm s}\doteq \left[\frac{d n_s}{d\log k}\right]_{k=k_*} \quad 
\beta_{\rm s}\doteq \left[\frac{d \alpha_s}{d\log k}\right]_{k=k_*}
\end{equation}
We adopt the same pivot scale of $k_{*}=0.05\,\rm{Mpc}^{-1}$ both for scalar and tensor perturbations. Notice that the running $\alpha_{\rm s}$ quantifies the rate of change of $n_{\rm s}$ per Hubble time (we recall that  $d/d\log k = 1/H\, d/dt$) while the running of running  $\beta_{\rm s}$ quantifies the rate of change of $\alpha_{\rm s}$ per Hubble time. These quantities are related to the shape of the inflationary potential (or equivalently to the dynamics of the background evolution) and consequently to the underlying physics of inflation.
Under the slow-roll assumption, $\alpha_s$ and $\beta_s$ can be both expressed in terms of the potential slow-roll parameters $\{\epsilon_{V}\,,\,\eta_V\,,\,\xi_V^2\,,\,\varpi_V^3\}$ as
\begin{equation}
\alpha_{\rm s} = + 16 \epsilon _ { V } \eta _ { V } - 24 \epsilon _ { V } ^ { 2 } - 2 \xi _ { V } ^ { 2 }    
\label{alpha_SR}
\end{equation}
\begin{align}
 \nonumber \beta_{\rm s} =&  - 192 \epsilon _ { V } ^ { 3 } + 192 \epsilon _ { V } ^ { 2 } \eta _ { V } - 32 \epsilon _ { V } \eta _ { V } ^ { 2 } - 24 \epsilon _ { V } \xi _ { V } ^ { 2 } \\ & + 2 \eta _ { V } \xi _ { V } ^ { 2 } + 2 \varpi _ { V } ^ { 3 }
 \label{beta_SR}
 \end{align}
or, equivalently, in terms of the parameters $\{\epsilon_i\}$ as 
\begin{equation}
\alpha_{\rm s} =-2\epsilon_1\epsilon_2-\epsilon_2\epsilon_3
\label{alpha_s}
\end{equation}
\begin{align}
\beta_{\rm s}=-2\epsilon_1 \epsilon_2^2 -2\epsilon_1\epsilon_2\epsilon_3-\epsilon_2\epsilon_3^2-\epsilon_2\epsilon_3\epsilon_4
 \end{align}
with $\epsilon_V\simeq \epsilon_1\simeq r/16$ which is clearly related to the amplitude of the tensor spectrum.

Similarly, for the tensor spectrum we adopt the parametrization 
\begin{align}
\nonumber \log\mathcal{P}_{\rm T}(k)=&\log\left(r\,A_{\mathrm{s}}\right)+\left(n_{\mathrm{T}}\right) \log\left(k / k_{*}\right) + \frac{\alpha_{\rm T}}{2} \log^2 \left(k / k_{*}\right)\\
& +\frac{\beta_{\rm T}}{6} \log^3 \left(k / k_{*}\right).
\end{align} 
We consider the tensor-to-scalar ratio $r\doteq A_T / A_s$  a free parameter while we use the slow-roll consistency relation $n_{\rm T}=-r/8$ for the tensor tilt. We also relate the higher order tensor runnings 
\begin{equation}
\alpha_{\rm T}\doteq \left[\frac{dn_{\rm T}}{d\log k}\right]_{k=k_*} \quad \beta_{\rm T}\doteq \left[\frac{d\alpha_{\rm T}}{d\log k}\right]_{k=k_*}
\end{equation}
to the scalar ones by a set of slow-roll consistency relations. Indeed, under the assumption of slow roll inflation, a set of consistency relations among scalar and tensor parameters can be derived at any order\footnote{It should be noted that these relations can be violated in many nonstandard inflationary models, e.g. in presence of other spectator (rolling) fields~\cite{Mukohyama:2014gba,Namba:2015gja,Peloso:2016gqs,Giare:2020vhn,Ozsoy:2020ccy} or in modified gravity theories~\cite{Baumann:2015xxa,Giovannini:2015kfa,Giovannini:2018dob,Giovannini:2018nkt,Giovannini:2018zbf,Giare:2020vss,Giare:2020plo,Cicoli:2020bao}.}~\cite{Martin:2013tda,Giare:2019snj}. In particular, the slow-roll consistency relations for the tensor running and running of running reads~\cite{Giare:2019snj}
\begin{equation}
\alpha_{\rm T}=\frac{r}{8}(n_s - 1) + \frac{r^2}{64},
\end{equation}
\begin{equation}
\beta_{\rm T}= \frac{r}{8}\left[ \alpha_{\rm s} - \left( n_{\rm s} -1 \right)^2\right] -\frac{3\,r^2}{64}\left(n_{\rm s} - 1\right) -\frac{r^3}{256}.
\end{equation} 
Therefore given constraint on the scalar spectral index $n_{\rm s}$, its running $\alpha_{\rm s}$ and on the tensor-to-scalar ratio $r$, constraints can be derived on the tensor spectral index $n_{\rm t}$, its running $\alpha_{\rm t}$, its running of running $\beta_{\rm t}$.

Finally, we consider also the curvature density parameter $\Omega_k$ as an additional free parameter of the cosmological model. We explore the possibility of a nontrivial background geometry as a consistency check of the standard slow-roll paradigm. Indeed the vast majority of inflationary models predict flatness and constraints on the spatial curvature are an important test of this standard scenario. 

\subsection{Numerical analyses and datasets}

\begin{table}
	\begin{center}
		\renewcommand{\arraystretch}{1.5}
		\begin{tabular}{c@{\hspace{0. cm}}@{\hspace{1.5 cm}} c}
			\hline
			\textbf{Parameter}    & \textbf{Prior} \\
			\hline\hline
			$\Omega_{\rm b} h^2$         & $[0.005\,,\,0.1]$ \\
			$\Omega_{\rm c} h^2$     	 & $[0.001\,,\,0.99]$\\
			$100\,\theta_{\rm {MC}}$     & $[0.5\,,\,10]$ \\
			$\tau$                       & $[0.01\,,\,0.8]$\\
			$\log(10^{10}A_{\rm S})$     & $[1.61\,,\,3.91]$ \\
			$n_{\rm s}$                  & $[0.8\,,\, 1.2]$ \\
			$\alpha_{\rm s}$             & $[-1\,,\, 1]$ \\ 
			$\beta_{\rm s}$              & $[-1\,,\, 1]$ \\
			$r$                          & $[0\,,\, 3]$ \\	
		    $\Omega_{\rm k}$             & $[-0.3\,,\,0.3]$\\
			\hline\hline
		\end{tabular}
		\caption{List of the parameter priors.}
		\label{tab.Priors}
	\end{center}
\end{table}
We perform Monte Carlo Markov Chain (MCMC) analyses using  the publicly available package \texttt{CosmoMC}~\cite{Lewis:2002ah,Lewis:2013hha} and computing the theoretical model described in the previous subsection with the latest version of the Boltzmann code \texttt{CAMB}~\cite{Lewis:1999bs,Howlett:2012mh}.
For all the different cosmological parameters we choose flat prior-distributions (unless otherwise stated), varying them uniformly in the conservative ranges listed in Table~\ref{tab.Priors}.
We explore the posteriors of our parameter space using the MCMC sampler developed for \texttt{CosmoMC} and tailored for parameter spaces with a speed hierarchy which also implements the "fast dragging" procedure described in Ref.~\cite{Neal:2005}. The convergence of the chains obtained with this procedure is tested using the Gelman-Rubin criterion~\cite{Gelman:1992zz} and we choose as a threshold for chain convergence $R-1 \lesssim 0.02 $.

Our baseline dataset consists of:
\begin{itemize}
	
	\item Planck 2018 temperature and polarization (TT TE EE) likelihood, which also includes low multipole data ($\ell < 30$)~\cite{Aghanim:2019ame,Aghanim:2018eyx,Akrami:2018vks}. We refer to this combination as "Planck."
	
	\item Planck 2018 lensing likelihood~\cite{Aghanim:2018oex}, constructed from measurements of the power spectrum of the lensing potential. We refer to this dataset as "lensing."

    \item Baryon acoustic oscillations (BAO) measurements extracted from data from the 6dFGS~\cite{Beutler:2011hx}, SDSS MGS~\cite{Ross:2014qpa} and BOSS DR12~\cite{Alam:2016hwk} surveys. We refer to this dataset combination as "BAO."

	\item CMB B-modes power spectrum likelihood cleaned from the foreground contamination as released by Bicep2/Keck Array X Collaboration \cite{Ade:2018gkx}. We refer to this dataset as "BK15."
	
	\item Atacama Cosmology Telescope DR4 likelihood, combined with WMAP 9-years observations data~\cite{Hinshaw:2012aka} and a Gaussian prior on $\tau = 0.065 \pm 0.015$, as done in~\cite{Aiola:2020azj}. We refer to this dataset combination as "ACTPol+WMAP."
	
	\item South Pole Telescope polarization measurements SPT-3G~\cite{Dutcher:2021vtw} combined with WMAP 9-years observations data~\cite{Hinshaw:2012aka} and a Gaussian prior on $\tau = 0.065 \pm 0.015$. We refer to this dataset combination as "SPT3G+WMAP."
	
\end{itemize}

We conclude this section with some important remarks about the different combinations of datasets that we use in the following discussion.

First of all we stress that for the Planck data we consider both the high-multipole likelihood (which includes multipoles $30\lesssim \ell \lesssim 2500$ for the TT spectrum and $30\lesssim \ell \lesssim 2000$ for TE and EE spectra) and the "low-E" polarization likelihood (which covers the multipole range $2\le \ell \le 30$ for the EE spectrum). In this way, analyzing the Planck anisotropies and polarization measurements, we can derive constraints on all the cosmological parameters of the model. Furthermore, we combine the Planck TT TE EE spectra with the Planck lensing measurement. Indeed the CMB photons that we measure today traversed almost the entire observable Universe and, along their paths, are deflected by gradients in the gravitational potentials associated with inhomogeneities in the Universe. This can cause a smoothing of the acoustic peaks and a conversion of E mode polarization into B-mode polarization. Therefore the Planck lensing reconstruction, being the most significant detection of CMB lensing to date, is useful to improve the constraints on cosmological parameters, providing sensitivity above all on parameters that affect the late-time expansion and the background geometry. However, while the Planck lensing measurements partially break the geometric degeneracy, it is well know that the inclusion of the baryon acoustic oscillation (BAO) measurements from galaxy surveys is a much more powerful way to break degeneracy in the geometrical sector. BAOs are the counterpart to the CMB acoustic peaks in the baryon distribution which remain imprinted also into the present-day matter distribution. Using the transverse BAOs information one can constrain the ratio between the comoving angular diameter distance ($D_M$) and the sound horizon ($r_d$) at the epoch when the baryon evolution becomes unaffected by coupling to photons. On the other hand, from the line-of-sight information we can constrain the quantity $H(z)\,r_d$. These two information can be combined together to constrain the acoustic-scale distance ratio $D_{V} / r_{\mathrm{d}} \doteq \left[c\, z\, D_{M}^{2}(z) H^{-1}(z)\right]^{1 / 3} / r_{\mathrm{d}}$. The acoustic scale measured by BAOs (at around $147$ Mpc), being much larger than the scale of virialized structures, makes the BAO measurements relatively simple geometric measurements insensitive to nonlinear physics, providing a robust geometrical test of cosmology. Here, in combination with the Planck data, we use the measurements of $D_{V} / r_{\mathrm{d}}$ from the 6dF survey at an effective redshift $z_{\rm eff} = 0.106$ \cite{Beutler:2011hx}, the SDSS Main Galaxy Sample at $z_{\rm eff}= 0.15$ \cite{Ross:2014qpa} and the final BOSS DR12 data with separate constraints on $H(z)\,r_{\rm d}$ and $D_M/r_{\rm d}$ in three correlated redshift bins at $z_{\rm eff} =\left[0.38\,,\,0.51\,,\,0.61\right]$ \cite{Alam:2016hwk}. Finally, to improve also the constraints in the primordial tensor sector, we exploit the CMB B-modes power spectrum likelihood (cleaned from the foreground contamination) as released by Bicep2/Keck Array X Collaboration \cite{Ade:2018gkx}. Indeed, it is well known that a satiable background of inflationary gravitational waves can produce B-modes polarization on large/intermediate angular scales where the cosmic variance is not very significant and gravitational lensing is not yet dominant. Notice however that the B-modes likelihood basically improves only the constraints on tensor modes. Therefore we include this dataset only when we analyze the tensor spectrum because interested in models with a satiable production of gravitational waves.

Along with these combinations of datasets involving the Planck CMB measurements, we analyze also two other Planck-independent datasets. In particular we use the Atacama Cosmology Telescope DR4 likelihood and the South Pole Telescope polarization measurements. We combine both of them with WMAP 9-years observations data~\cite{Hinshaw:2012aka}. The reason is that the Atacama Cosmology Telescope has a minimum sensitivity in multipole of $600$ in TT, and $350$ in TE and EE, and so it lacks data around the first two acoustic peaks in the TT spectrum and the first full peak in TE/EE. Similarly, the South Pole Telescope measures only the TE and EE spectra over a range of multipoles $300\le \ell \le 1400$ for EE and $300 \le \ell \le 1700$ for TE. Therefore, the only way to obtain competitive Planck-independent measurements for all the cosmological parameters is to combine these two datasets with the public WMAP 9-year observations at intermediate scales ($2<\ell< 1200$ for TT and $\ell < 800$ for TE), as also done in~\cite{Dutcher:2021vtw,Aiola:2020azj}. Notice also that we use a Gaussian prior on $\tau = 0.065 \pm 0.015$ both for ACT+WMAP and for SPT3G+WMAP. Indeed, while our primary goal is to obtain a measurement of the cosmological (inflationary) parameters that is Planck-independent, neither ACT nor SPT-3G can constrain the optical depth at reionization $\tau$. Furthermore there is evidence that WMAP large-scale polarization data ($2 < \ell < 23$ in TE spectrum) can be contaminated by dust, possibly affecting the WMAP bounds on $\tau$. For this reason in our analysis we exclude this multipoles range, using instead the conservative Gaussian prior on $\tau$ which is based on Planck measurements. This prior on $\tau$ is not expected to affect the constraints on the other cosmological parameters~\cite{Dutcher:2021vtw,Aiola:2020azj}.

\section{Results} \label{sec.Results}
In this section we present and discuss the results obtained by our MCMC analysis of the inflationary slow-roll paradigm in extended parameter spaces beyond the standard $\Lambda\rm{CDM}$ cosmological model.


\begin{table*}
	\begin{center}
		\renewcommand{\arraystretch}{1.5}
		\resizebox{0.9\textwidth}{!}{\begin{tabular}{c c c c  c  c}
  	        \hline
			\textbf{Parameter} & \textbf{Planck18}  & \textbf{Planck18 + lensing}  & \textbf{Planck18 + BAO} & \textbf{ACTPol + WMAP} & \textbf{SPT3G+WMAP} \\
			\hline\hline
			$\Omega_{\rm b} h^2$ &$0.02235\pm 0.00017$&$0.02237\pm 0.00016$&$0.02243\pm 0.00015$ & $0.02195\pm 0.00025$ &$0.02251\pm 0.00025$\\
			$\Omega_{\rm c} h^2$ &$0.1207\pm 0.0015$&$0.1202\pm 0.0012$&$0.1195\pm 0.0010$ & $0.1190\pm 0.0029$ &$0.1139\pm 0.0032$\\
			$100\,\theta_{\rm {MC}}$ &$1.04085\pm 0.00031$&$1.04089\pm 0.00030$&$1.04100\pm 0.00028$& $1.04174\pm 0.00066$ &$1.03970\pm 0.00066$\\
			$\tau$   &$0.0575\pm 0.0086$&$0.0564\pm 0.0080$&$0.0590\pm 0.0087$& $0.061\pm 0.013$ &$0.063\pm 0.013$\\
			$\log(10^{10}A_{\rm S})$ &$3.053\pm 0.018$&$3.049\pm 0.015$&$3.053\pm 0.018$ & $3.051\pm 0.026$ &$3.037\pm 0.026$\\
			$n_s$ &$0.9612\pm 0.0054$&$0.9625\pm 0.0048$&$0.9645\pm 0.0045$ & $0.9680\pm 0.0082$ &$0.978\pm 0.011$\\
			$\alpha_s$ &$0.001\pm 0.010$&$0.002\pm 0.010$&$0.000\pm 0.010$ & $0.035\pm 0.012$ &$0.028\pm 0.017$\\
			$\beta_s$ &$0.012\pm 0.013$&$0.010\pm 0.013$&$0.009\pm 0.013$ & $0.035\pm 0.013$ &$0.023\pm 0.016$\\
			\hline 
			$\eta_V$ &$-0.0194^{+0.0027}_{-0.0026}$&$-0.0187^{+0.0025}_{-0.0023}$&$-0.0177^{+0.0021}_{-0.0022}$ & $-0.0160\pm 0.0041$ &$-0.0111\pm 0.0053$\\
			$\xi^2_V$ &$-0.0005\pm0.0050$&$-0.0008^{+0.0050}_{-0.0049}$&$-0.0001\pm0.0049$ & $-0.0174\pm 0.0058$ &$-0.0141\pm 0.0085$\\
			$\varpi _ { V } ^ { 3 }$ &$0.0058^{+0.0063}_{-0.0061}$&$0.0051^{+0.0062}_{-0.0061}$&$0.0046^{+0.0062}_{-0.0061}$ & $0.0172\pm 0.0064$ &$0.0115\pm 0.0078$\\
			$\epsilon_2$ &$0.0388^{+0.0053}_{-0.0054}$&$0.0375^{+0.0047}_{-0.0049}$&$0.0355^{+0.0044}_{-0.0043}$ & $0.0320\pm 0.0082$ &$0.022\pm 0.011$\\
			$\epsilon_3$ &$-0.02\pm0.26$&$-0.04^{+0.27}_{-0.26}$&$0.00\pm0.28$ & $<-0.02$ &$-$\\
			\hline	\hline
		\end{tabular}}
	\end{center}
	\caption{Results for $\Lambda\rm CDM + \alpha_s + \beta_s$. The constraints on parameters are at $68\%$ CL, while upper bounds are at $95\%$ CL. The internal horizontal line divides the primary parameters of the cosmological model (those we directly sample in our MCMC analysis) from the derived parameters (those we obtain from the others by the relations described in the text).}
	\label{tab.LCDM+runnings}
\end{table*}

\begin{figure*}
	\centering
	\includegraphics[width=0.7 \textwidth]{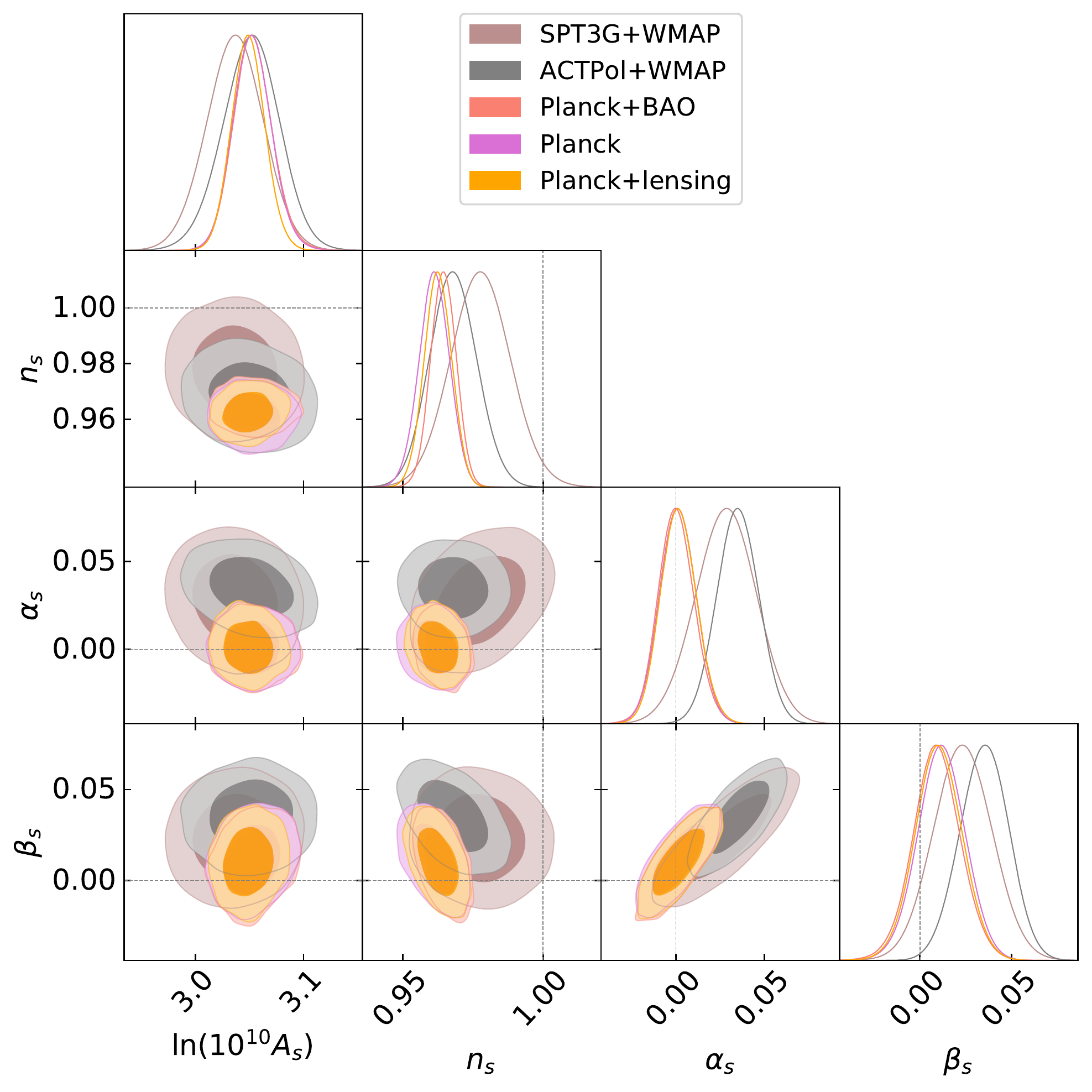}
	\caption{Marginalized 2D and 1D posteriors distributions for the $\Lambda\rm CDM + \alpha_s + \beta_s$ cosmological model obtained for the different combinations of the datasets listed in Sec.~\ref{sec.Methods}. The dashed lines represent the case of vanishing inflationary parameters.}
	\label{fig:LCDM+runnings}
\end{figure*}

\begin{table*}
	\begin{center}
		\renewcommand{\arraystretch}{1.5}
		\resizebox{\textwidth}{!}{\begin{tabular}{c c c c c c c}
			\hline
			\textbf{Parameter} & \textbf{Planck18}  & \textbf{Planck18 + lensing}  & \textbf{Planck18 + BAO} & \textbf{ Planck18 + BK15} & \textbf{ACTPol + WMAP} & \textbf{SPT3G+WMAP} \\
			\hline\hline
			$\Omega_{\rm b} h^2$ &$0.02241\pm 0.00016$&$0.02242\pm 0.00015$&$0.02247\pm 0.00014$&$0.02239\pm 0.00015$ & $0.02234\pm 0.00022$ &$0.02273\pm 0.00024$\\
			$\Omega_{\rm c} h^2$ &$0.1202\pm 0.0014$&$0.1199\pm 0.0012$&$0.1193\pm 0.0010$&$0.1206\pm 0.0014$ & $0.1179\pm 0.0030$ &$0.1138\pm 0.0031$\\
			$100\,\theta_{\rm {MC}}$ &$1.04091\pm 0.00032$&$1.04093\pm 0.00030$&$1.04101\pm 0.00030$&$1.04087\pm 0.00031$ & $1.04186\pm 0.00065$ &$1.03978\pm 0.00067$\\
			$\tau$   &$0.0562\pm 0.0081$&$0.0560\pm 0.0076$&$0.0573\pm 0.0080$&$0.0570\pm 0.0083$& $0.058\pm 0.012$ &$0.060\pm 0.013$\\
			$\log(10^{10}A_{s})$ &$3.050\pm 0.017$&$3.049\pm 0.015$&$3.051\pm 0.017$&$3.053\pm 0.017$ & $3.049\pm 0.025$  &$3.037\pm 0.026$\\
			$n_s$ &$0.9642\pm 0.0047$&$0.9647\pm 0.0044$&$0.9665\pm 0.0041$&$0.9629\pm 0.0046$ & $0.9796\pm 0.0074$ &$0.980\pm 0.010$\\
			$\alpha_s$  &$-0.0094\pm 0.0074$&$-0.0084\pm 0.0073$&$-0.0091\pm 0.0075$&$-0.0080\pm 0.0069$ & $ 0.0090\pm 0.0087$  &$0.001\pm 0.012$\\
			$r$ &$<0.165 $&$<0.159$&$< 0.172$&$< 0.0658$ & $<0.176$ &$<0.260$\\
			\hline
			$n_T$ &$>-0.0206$&$>-0.0198$&$>-0.0215$&$>-0.0082$ & $>-0.022$ &$>-0.032$\\
			$\alpha_T$ &$\left(\,-18^{+12}_{-10}\,\right)\cdot 10^{-5}$&$\left(-17\pm 10\right)\cdot 10^{-5}$&$\left(\,-16.6^{+11}_{-9.5}\,\right)\cdot 10^{-5}$&$\left(\,-11.7^{+7.9}_{-5.9}\,\right)\cdot 10^{-5}$ & $\left(\,-4.2^{+6.9}_{-10}\,\right)\cdot 10^{-5}$  &$\left(\,3^{+13}_{-27}\,\right)\cdot 10^{-5}$\\
			$\beta_T$ &$\left(11.4^{+6.9}_{-15}\right)\cdot 10^{-5}$&$\left(9.96^{+6.1}_{-14}\right)\cdot 10^{-5}$&$\left(11.8^{+7.2}_{-16}\right)\cdot 10^{-5}$ &$\left(\,3.9^{+2.5}_{-4.8}\,\right)\cdot 10^{-5}$ & $\left(\,-4.4^{+8.1}_{-6.9}\,\right)\cdot 10^{-5}$  &$\left(\,5^{+12}_{-21}\,\right)\cdot 10^{-5}$\\
			$\epsilon_V\simeq\epsilon_1$ &$< 0.0103$&$< 0.0099 $&$<0.0108$&$< 0.0041$ & $< 0.0110$ &$< 0.0163$\\
			$\eta_V$ &$-0.0058^{+0.0069}_{-0.012}$&$-0.0061^{+0.0066}_{-0.011}$&$-0.0039^{+0.0072}_{-0.012}$&$-0.0130^{+0.0038}_{-0.0050}$ & $0.0015^{+0.0074}_{-0.013}$  &$0.010^{+0.012}_{-0.019}$\\
			$\xi_V^2$ &$0.0044\pm 0.0037$&$0.0040\pm 0.0036$&$0.0043\pm 0.0038$&$0.0038\pm 0.0034$& $-0.0045\pm 0.0044$  &$-0.0001^{+0.0056}_{-0.0064}$\\
			$\epsilon_2$ &$0.0277^{+0.0095}_{-0.0067}$&$0.0276^{+0.0091}_{-0.0062}$&$0.0250^{+0.0095}_{-0.0064}$&$0.0334\pm 0.0054$ & $ 0.0126^{+0.012}_{-0.0090}$ &$0.006^{+0.016}_{-0.013}$\\
			$\epsilon_3$ &$-$&$0.37^{+0.26}_{-0.34}$&$0.62^{+0.16}_{-0.56}$&$0.24\pm 0.21$ & $-$ &$-$\\
			$V_{\rm inf}^{1/4}$ & $<2.04\times10^{16}\,\rm GeV$&$<2.01\times10^{16}\,\rm GeV$&$<2.06\times10^{16}\,\rm GeV$&$<1.62\times10^{16}\,\rm GeV$ & $<2.10\times10^{16}\,\rm GeV$ &$<2.31\times10^{16}\,\rm GeV$ \\
			\hline	\hline
		\end{tabular}}
	\end{center}
	\caption{Results for $\Lambda\rm CDM + r+ \alpha_s$. The constraints on parameters are at $68\%$ CL, while upper bounds are at $95\%$ CL. The internal horizontal line divides the primary parameters of the cosmological model (those we directly sample in our MCMC analysis) from the derived parameters (those we obtain from the others by the relations described in the text).}
	\label{tab.LCDM+r+nrun}
\end{table*}

\begin{figure*}
	\centering
	\includegraphics[width=1 \textwidth]{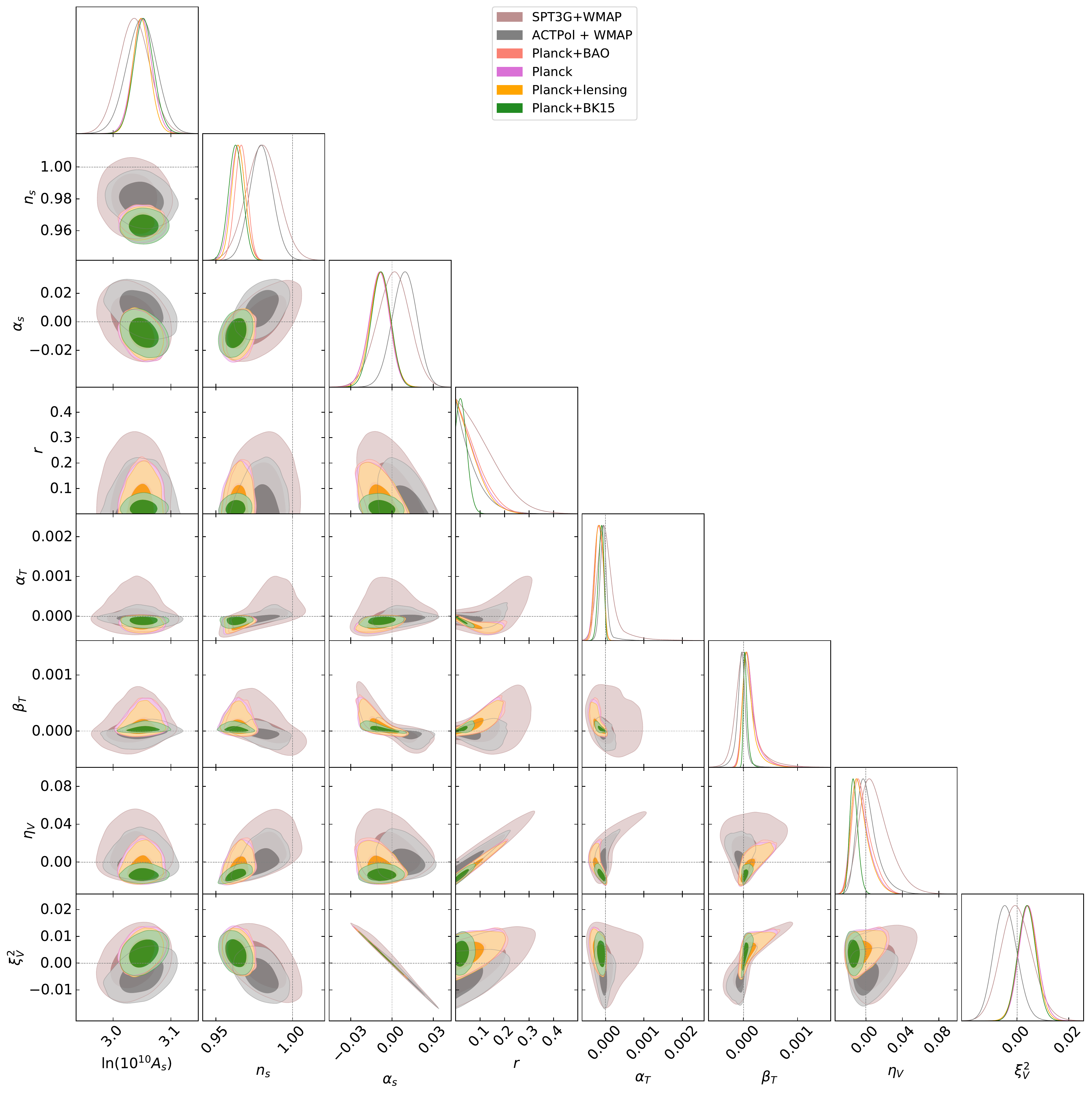}
	\caption{Marginalized 2D and 1D posteriors distributions for the $\Lambda\rm CDM + r+ \alpha_s$ cosmological model obtained for different combination of datasets listed in Sec.~\ref{sec.Methods}. The dashed lines represent the case of vanishing inflationary parameters.}
	\label{fig:LCDM+r+nrun}
\end{figure*}

\begin{figure*}
	\centering
	\includegraphics[width=1 \textwidth]{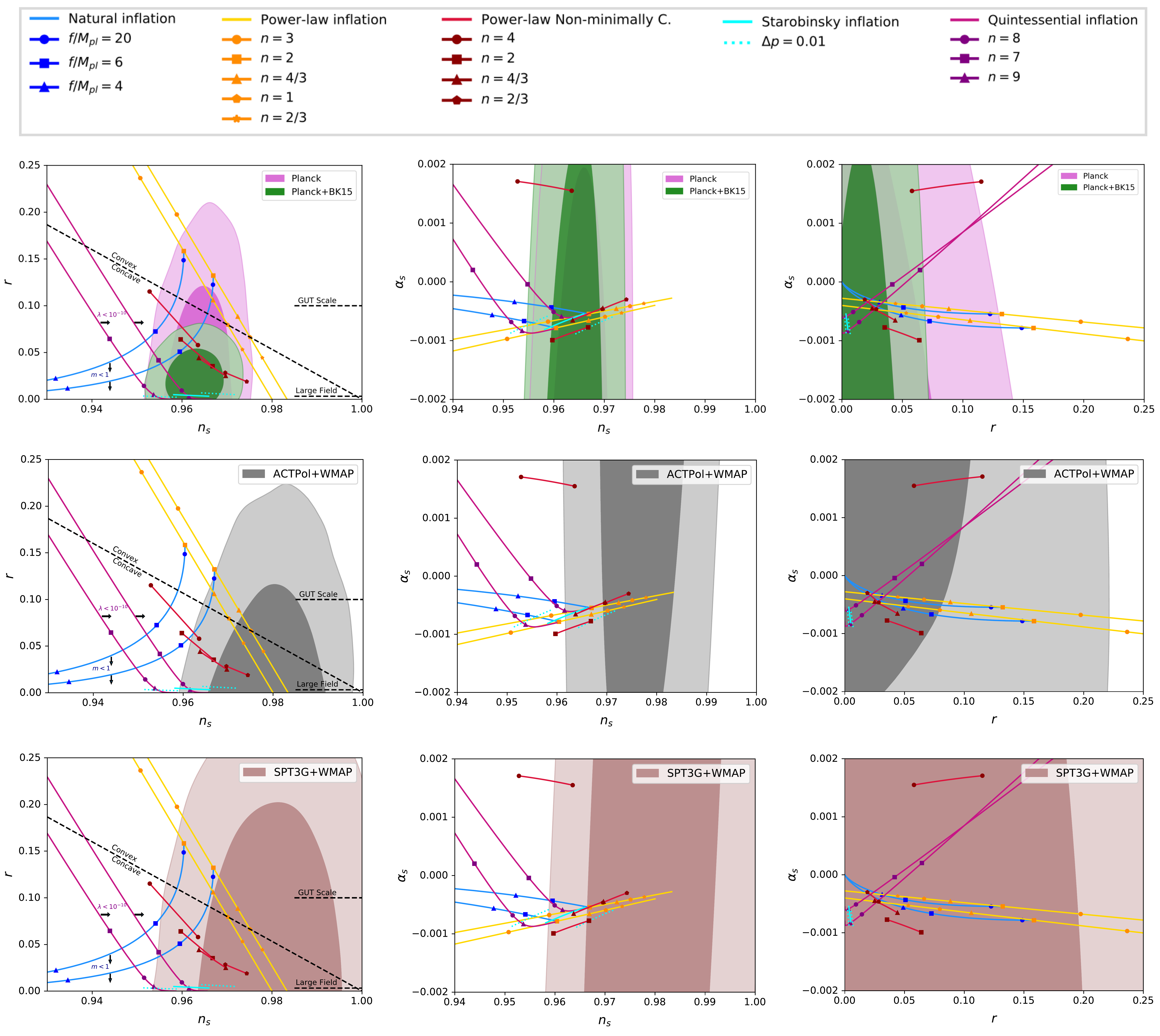}
	\caption{Marginalized joint 68\% and 95\% CL regions for $(n_s\,,\,r)$ , $(n_s\,,\,\alpha_s)$ and  $(r\,,\,\alpha_s)$ from Planck(+BK15) (top panels), ACTPol+WMAP (middle panels) and SPT3G+WMAP (bottom panels) data. The marginalized contours can be compared to the theoretical predictions of some selected inflationary models opportunely described in the text.}
	\label{fig:Models}
\end{figure*}

\begin{table*}
	\begin{center}
		\renewcommand{\arraystretch}{1.5}
		\resizebox{\textwidth}{!}{\begin{tabular}{c c c c c c c}
			\hline
			\textbf{Parameter} & \textbf{Planck18}  & \textbf{Planck18 + lensing}  & \textbf{Planck18 + BAO} & \textbf{Planck18 + BK15} & \textbf{ACTPol + WMAP} & \textbf{SPT3G+WMAP} \\
			\hline\hline
			$\Omega_{\rm b} h^2$ &$0.02263\pm 0.00018$&$0.02252\pm 0.00017$&$0.02241\pm 0.00015$&$0.02262\pm 0.00017$  &$0.02245\pm 0.00022$ &$0.02273\pm 0.00025$\\
			$\Omega_{\rm c} h^2$ &$0.1177\pm 0.0016$&$0.1181\pm 0.0015$&$0.1196\pm 0.0014$&$0.1179\pm 0.0015$ &$0.1184\pm 0.0030$ &$0.1141\pm 0.0033$\\
			$100\,\theta_{\rm {MC}}$ &$1.04120\pm 0.00033$&$1.04110\pm 0.00032$&$1.04097\pm 0.00031$&$1.04118\pm 0.00033$ &$1.04181\pm 0.00065$ &$1.03975\pm 0.00070$\\
			$\tau$   &$0.0480^{+0.0087}_{-0.0072}$&$0.0487^{+0.0085}_{-0.0075}$&$0.0554\pm 0.0080$&$0.0477^{+0.0086}_{-0.0072}$ &$0.059\pm 0.013$ &$0.060\pm 0.013$\\
			$\log(10^{10}A_{s})$ &$3.026^{+0.018}_{-0.015}$&$3.027^{+0.018}_{-0.016}$&$3.045^{+0.015}_{-0.017}$&$3.026\pm 0.018$ &$3.057\pm 0.027$ &$3.039\pm 0.026$ \\
			$n_s$ &$0.9728\pm 0.0052$&$0.9707\pm 0.0049$&$0.9671\pm 0.0046$&$0.9715\pm 0.0048$ &$0.9773\pm 0.0070$ &$0.9793\pm 0.0091$\\
			$r$ &$<0.170$&$<0.154$&$<0.120$&$<0.0613$ &$< 0.210$ &$< 0.259$\\
			$\Omega_k$&$-0.048^{+0.020}_{-0.016}$&$-0.0123^{+0.0072}_{-0.0063}$&$0.0007\pm 0.0020$&$-0.047^{+0.018}_{-0.015}$ &$-0.007^{+0.016}_{-0.012}$ &$0.0008^{+0.013}_{-0.0097}$\\
			\hline
			$n_T$ &$>-0.0212$&$>-0.0192$&$>-0.0150$&$>-0.0077$ &$>-0.0262$ &$>-0.0324$\\
			$\alpha_T$ &$\left(\,-10.8\pm 8.5\,\right)\cdot 10^{-5}$&$\left(-12\pm 7.8\right)\cdot 10^{-5}$&$\left(\,-12.7^{+9.5}_{-7.3}\,\right)\cdot 10^{-5}$&$\left(\,-7.5^{+5.6}_{-3.8}\,\right)\cdot 10^{-5}$ &$\left(\,-3.7^{+8}_{-16}\,\right)\cdot 10^{-5}$ &$\left(\,5^{+14}_{-31}\,\right)\cdot 10^{-5}$\\
			$\epsilon_V\simeq\epsilon_1$ &$<0.0106$&$<0.0097$&$<0.0075$&$<0.0038$ &$< 0.0131$ &$< 0.0162$\\
			$\eta_V$ &$-0.0005^{+0.0081}_{-0.013}$&$-0.0033^{+0.0069}_{-0.012}$&$-0.0079^{+0.0053}_{-0.0091}$&$-0.0094^{+0.0038}_{-0.0049}$ &$0.005^{+0.010}_{-0.016}$ &$ 0.0096^{+0.013}_{-0.021}$\\
			$\epsilon_2$ &$0.0184^{+0.011}_{-0.0080}$&$0.0217^{+0.0098}_{-0.0070}$&$0.0272^{+0.0081}_{-0.0058}$&$0.0252\pm 0.0055$ &$0.012^{+0.015}_{-0.010}$ &$0.007^{+0.019}_{-0.013}$\\
			$V_{\rm inf}^{1/4}$ & $<2.08\times10^{16}\,\rm GeV$&$<2.03\times10^{16}\,\rm GeV$&$<1.90\times10^{16}\,\rm GeV$&$<1.61\times10^{16}\,\rm GeV$ &$<2.19\times10^{16}\,\rm GeV$ &$<2.31\times10^{16}\,\rm GeV$\\
			$\Delta N_{\rm tot}$ &$63.55^{+0.30}_{-0.21}$&$-$&$-$&$63.31^{+0.31}_{-0.23}$ &$-$ &$-$\\
			$\Delta N(k_{\rm exit})$ &$1.55^{+0.30}_{-0.21}$&$-$&$-$&$1.31^{+0.31}_{-0.23}$ &$-$ &$-$\\
			\hline	\hline
		\end{tabular}}
	\end{center}
	\caption{Results for $\Lambda\rm CDM + r+ \Omega_k$. The constraints on parameters are at $68\%$ CL, while upper bounds are at $95\%$ CL. The internal horizontal line divides the primary parameters of the cosmological model (those we directly sample in our MCMC analysis) from the derived parameters (those we obtain from the others by the relations described in the text).}
	\label{tab.LCDM+r+omegak}
\end{table*}

\begin{figure*}
	\centering
	\includegraphics[width=0.8\textwidth]{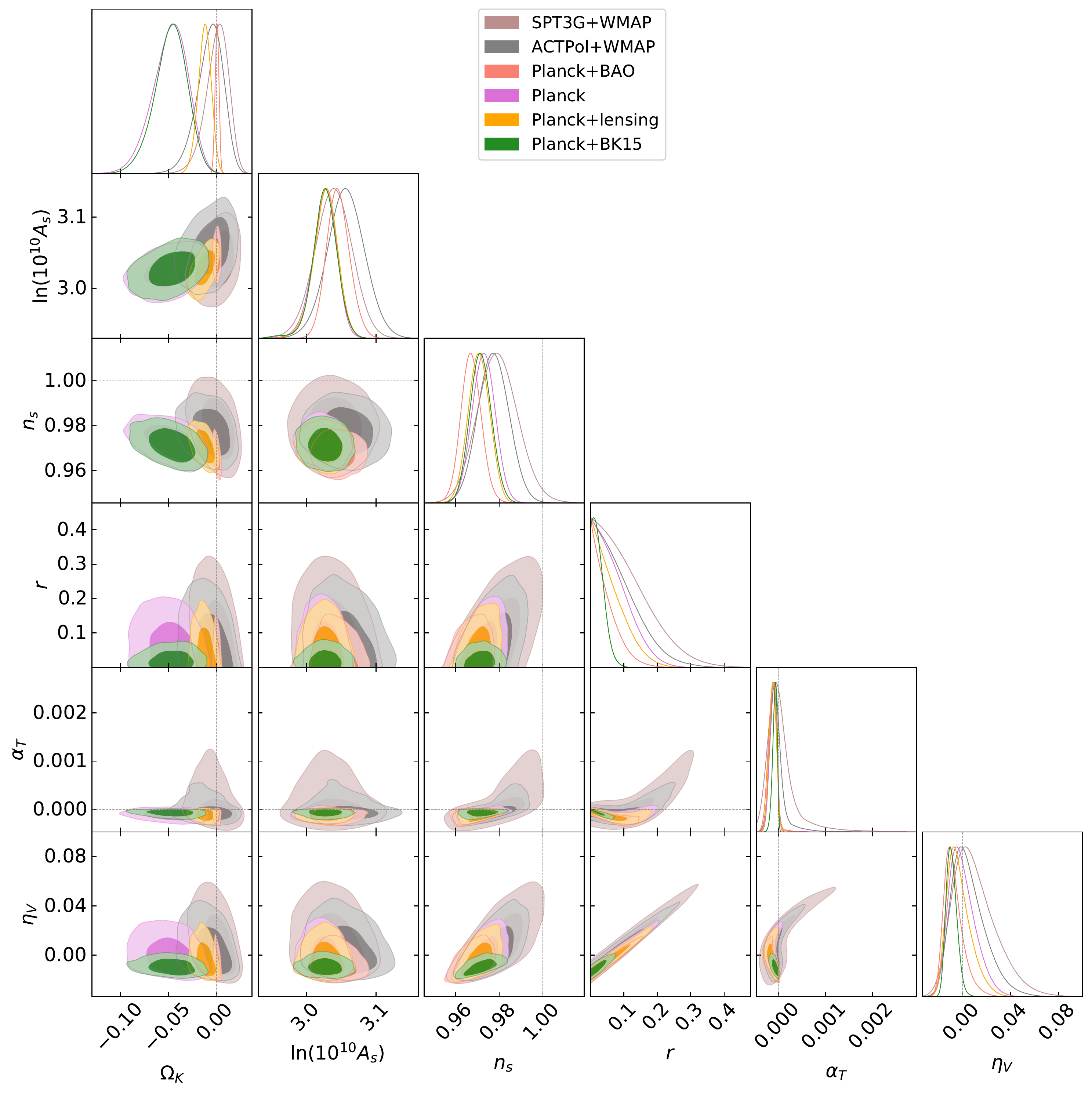}
	\caption{Marginalized 2D and 1D posteriors distributions for the $\Lambda\rm CDM + r+ \Omega_k$ cosmological model obtained for different combinations of the datasets listed in Sec.~\ref{sec.Methods}. The dashed lines represent the  case of vanishing inflationary parameters and flat spacetime geometry.}
	\label{fig:LCDM+r+omegak}
\end{figure*}

\begin{table*}
	\begin{center}
		\renewcommand{\arraystretch}{1.5}
		\resizebox{\textwidth}{!}{\begin{tabular}{c c c c c c c}
			\hline
			\textbf{Parameter} & \textbf{Planck18}  & \textbf{Planck18 + lensing}  & \textbf{Planck18 + BAO} & \textbf{Planck18 + BK15} & \textbf{ACTPol + WMAP} & \textbf{SPT3G+WMAP}\\
			\hline\hline
			$\Omega_{\rm b} h^2$ &$0.02268\pm0.00018$&$0.02255\pm0.00017$&$0.02245\pm0.0016$&$0.02263\pm0.00017$ &$0.02236\pm 0.00022$ &$0.02274\pm 0.00024$\\
			$\Omega_{\rm c} h^2$ &$0.1176\pm0.0016$&$0.1182\pm0.0015$&$0.1197\pm0.0015$&$0.1180\pm0.0015$ &$0.1171\pm 0.0032$ &$0.1141\pm 0.0038$\\
			$100\,\theta_{\rm {MC}}$ &$1.04121\pm0.00033$&$1.04110\pm0.00032$&$1.04097\pm0.00032$&$1.04118\pm0.00032$ &$1.04189\pm 0.00067$ &$1.03979\pm 0.00069$\\
			$\tau$   &$0.0491\pm0.0085$&$0.0514\pm0.0083$&$0.0573^{+0.0077}_{-0.0086}$&$0.0487\pm0.0086$ &$0.056^{+0.013}_{-0.012}$ &$0.060\pm 0.013$\\
			$\log(10^{10}A_{s})$ &$3.029\pm0.018$&$3.034\pm0.018$&$3.052\pm0.018$&$3.029\pm0.018$  &$3.043\pm 0.028$ &$3.038\pm 0.029$\\
			$n_s$ &$0.9720\pm0.0052$&$0.9696\pm0.0051$&$0.9655\pm0.0048$&$0.9710$ &$0.9810\pm 0.0077$ &$0.980\pm 0.012$\\
			$\alpha_s$  &$-0.0078\pm0.0080$&$-0.0064^{+0.0078}_{-0.0070}$&$-0.0097\pm0.0076$&$-0.0029\pm0.0068$ &$0.0102\pm 0.0090$ &$0.000\pm 0.013$\\
			$r$ &$<0.250$&$<0.205$&$<0.188$&$<0.0637$ &$<0.185$ &$< 0.282$\\
			$\Omega_k$&$-0.048^{+0.020}_{-0.016}$&$-0.0113\pm0.0066$&$0.0007\pm0.0020$&$-0.046^{+0.017}_{-0.014}$ &$-0.010^{+0.017}_{-0.011}$ &$0.000^{+0.015}_{-0.011}$\\
			\hline
			$n_T$ &$>-0.0312$&$>-0.0256$&$>-0.0235$&$>-0.0080$ &$>-0.0231$ &$>-0.0352$\\
			$\alpha_T$ &$\left(\,-9.6^{+8.4}_{-15}\,\right)\cdot 10^{-5}$&$\left(-13.6^{+8.8}_{-10}\right)\cdot 10^{-5}$&$\left(\,-17\pm 11\,\right)\cdot 10^{-5}$&$\left(\,-7.9^{+6.0}_{-3.9}\,\right)\cdot 10^{-5}$ &$\left(\,-2.5^{+2.7}_{-8.8}\,\right)\cdot 10^{-5}$ &$\left(\,5^{+16}_{-34}\,\right)\cdot 10^{-5}$\\
			$\beta_T$ &$\left(16^{+11}_{-22}\right)\cdot 10^{-5}$&$\left(10.6^{+7.3}_{-16}\right)\cdot 10^{-5}$&$\left(13.4^{+8}_{-18}\right)\cdot 10^{-5}$ &$\left(\,1.7^{+2.0}_{-3.4}\,\right)\cdot 10^{-5}$ &$\left(\,-5.8^{+9.3}_{-7.3}\,\right)\cdot 10^{-5}$ &$\left(\,6^{+16}_{-25}\,\right)\cdot 10^{-5}$\\
			$\epsilon_V\simeq\epsilon_1$ &$<0.0156$&$< 0.0128 $&$<0.0118$&$<0.0040$ &$< 0.0116$ &$< 0.0176$\\
			$\eta_V$ &$0.006^{+0.011}_{-0.018}$&$0.0003^{+0.0089}_{-0.015}$&$-0.0034^{+0.0079}_{-0.014}$&$-0.0095^{+0.0037}_{-0.0049}$ &$0.0030^{+0.0079}_{-0.014}$ &$0.011^{+0.013}_{-0.021}$\\
			$\xi_V^2$ &$0.0040^{+0.0039}_{-0.0044}$&$0.0031^{+0.0035}_{-0.0040}$&$0.0046\pm 0.0038$&$0.0013\pm 0.0034$ &$-0.0050\pm 0.0046$ &$0.0004\pm0.0066$\\
			$\epsilon_2$ &$0.0146^{+0.014}_{-0.0096}$&$0.0201^{+0.012}_{-0.0082}$&$0.0253^{+0.011}_{-0.0074}$&$ 0.0256\pm 0.0057$ &$0.0107^{+0.013}_{-0.0095}$ &$0.006^{+0.019}_{-0.016}$\\
			$\epsilon_3$ &$-$&$-$&$-$&$0.10\pm 0.29$ &$-$ &$-$\\
			$V_{\rm inf}^{1/4}$ & $<2.3\times10^{16}\,\rm GeV$&$<2.2\times10^{16}\,\rm GeV$&$<2.1\times10^{16}\,\rm GeV$&$<1.6\times10^{16}\,\rm GeV$ &$<2.12\times10^{16}\,\rm GeV$ &$<2.35\times10^{16}\,\rm GeV$\\
			$\Delta N_{\rm tot}$ &$63.67^{+0.29}_{-0.21}$&$-$&$-$&$63.33^{+0.30}_{-0.22}$ &$-$ &$-$\\
			$\Delta N(k_{\rm exit})$ &$ 1.67^{+0.29}_{-0.21}$&$-$&$-$&$1.34^{+0.30}_{-0.22}$ &$-$ &$-$\\
			\hline	\hline
		\end{tabular}}
	\end{center}
	\caption{Results for $\Lambda\rm CDM + r+ \alpha_s+ \Omega_k$. The constraints on parameters are at $68\%$ CL, while upper bounds are at $95\%$ CL.The internal horizontal line divides the primary parameters of the cosmological model (those we directly sample in our MCMC analysis) from the derived parameters (those we obtain from the others by the relations described in the text).}
	\label{tab.LCDM+r+nrun+omegak}
\end{table*}

\begin{figure*}
	\centering
	\includegraphics[width=1 \textwidth]{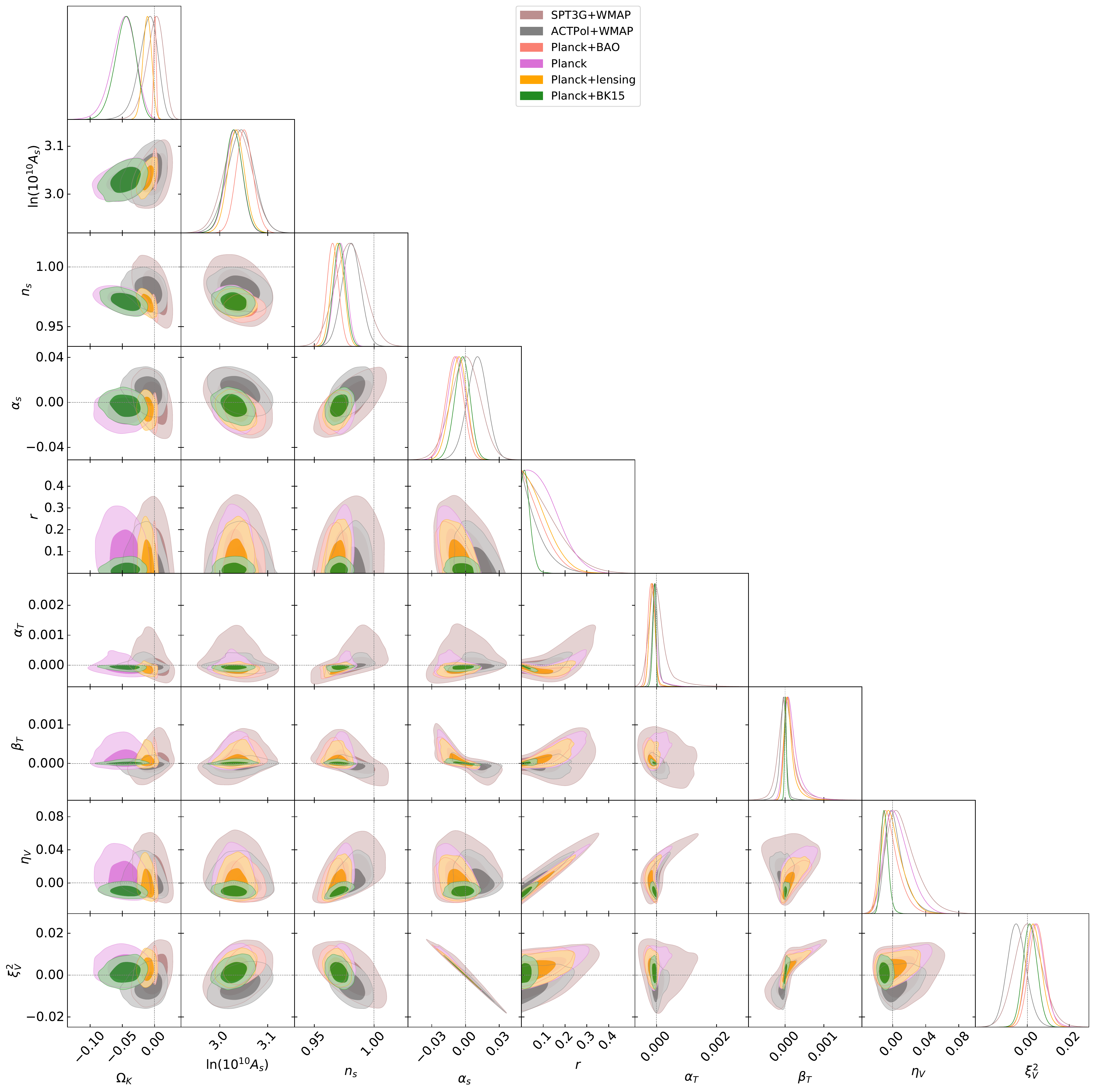}
	\caption{Marginalized 2D and 1D posteriors distributions for the $\Lambda\rm CDM + r+ \alpha_s+\Omega_k$ cosmological model obtained for different combinations of the datasets listed in Sec.~\ref{sec.Methods}. The dashed lines represent the  case of vanishing inflationary parameters and flat spacetime geometry.}
	\label{fig:LCDM+r+nrun+omegak}
\end{figure*}

\begin{figure*}
	\centering
	\includegraphics[width=0.85 \textwidth]{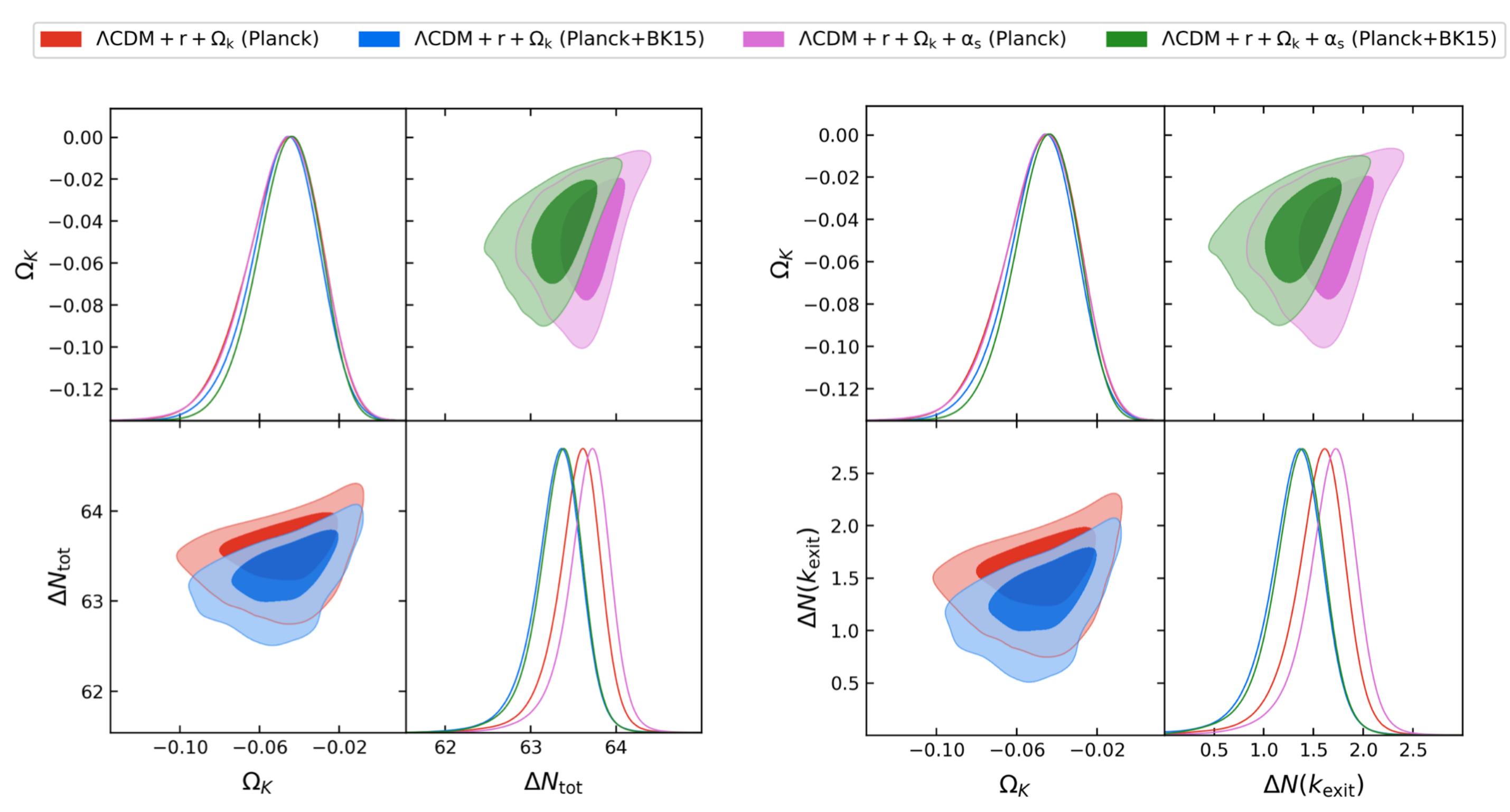}
	\caption{Marginalized 2D and 1D posteriors for the total number of \textit{e}-fold of inflation $\Delta N_{\rm tot}$ in a closed cosmological spacetime (left panel) and for the number of \textit{e}-fold before the largest observable scale exits the horizon during inflation $\Delta N(k_{\rm exit})$ (right panel).}
	\label{fig:Ntot_exit}
\end{figure*}


\subsection{Running the scalar running}

We start analyzing an extended cosmological model which includes both the running of the scalar spectral index $\alpha_s\doteq dn_s/d\log k$ and its running of running $\beta_{\rm s}\doteq d\alpha_s/d\log k$ as additional parameters. We refer to this model as $\Lambda\rm{CDM}+\alpha_s+\beta_s$. Notice that here we focus exclusively on the adiabatic scalar modes, parametrizing the scalar spectrum by Eq.\eqref{PhenomenologicalScalar} and assuming a negligible gravitational waves production\footnote{Assuming a negligible tensor amplitude $r=16\epsilon_V\simeq 16\epsilon_1\sim 0$ in terms of the slow-roll parameter means to consider $\epsilon_{V} \simeq \epsilon_1\sim 0$; e.i., negligibly small in the relations for the scalar tilt and its runnings.}.

In Table~\ref{tab.LCDM+runnings} we summarize the results obtained for this model while in Fig.~\ref{fig:LCDM+runnings} we show the $68\%$ and $95\%$~CL contour plots for different parameters. 

From the Planck data we derive the constraints on the scalar tilt $n_s=0.9612\pm 0.0054$, on its running $\alpha_s=0.001\pm 0.010$, and on its running of running $\beta_s=0.012\pm 0.013$, all at 68\% CL\footnote{Unless otherwise stated, we always provide 68\% CL values for bounded parameters and $95\%$ CL for upper/lower bounds.}. The inclusion of the lensing spectrum and the BAO data does not change significantly the above mentioned constraints
and all these bounds are consistent with the case of vanishing runnings within one standard deviation, see also Fig.\ref{fig:LCDM+runnings}. We compare the Planck results with other independent measurements derived using the different datasets listed in Sec.~\ref{sec.Methods}. Considering the SPT-3G data combined with WMAP 9-years observations data, we get $\alpha_s=0.028\pm 0.017$ and $\beta_s=0.023\pm 0.016$, both and consistent with zero within 1.6 and 1.4 standard deviations, respectively. On the other hand, considering the ACTPol+WMAP data, we obtain a preference for nonvanishing running $\alpha_s=0.035\pm0.012$ and for a nonvanishing running of running $\beta_s=0.035\pm0.013$ at the level of 2.9$\sigma$ and 2.7$\sigma$, respectively. Interestingly, in both the cases positive values for the runnings are preferred with a statistical significance between about $1.7\,\sigma$ (SPT-3G+WMAP) and $2.9\,\sigma$ (ACTPol+WMAP). Notice also that, 
while both the ground based telescope measurements are in good agreement one with each other, they are in disagreement at more than $2\sigma$ with Planck regarding the value of the running $\alpha_s$, and in tension for the running of running $\beta_s$ (see also Fig.~\ref{fig:LCDM+runnings}). This tension indicates a difference coming from the high multipoles region, that can be an indication of small systematic errors unaccounted for, or physics beyond the standard models. In other words, the extended models considered in this paper recast the global tension between the datasets already present for a $\Lambda$CDM model~\citep{Handley:2020hdp} analysis.

Under the assumption of a negligible tensor amplitude, we derive constraints on the slow-roll parameters $\{\eta_V\,,\,\xi_V^2\,,\,\varpi_V^3\}$ and $\{\epsilon_{i=2,3,4}\}$.
Due to the Planck data evidence for a tilted scalar spectrum, we obtain nonzero slow-roll parameters $\eta_V=-0.0194^{+0.0027}_{-0.0026}$ or equivalently $\epsilon_2=0.0388^{+0.0053}_{-0.0054}$. On the other hand, the missing evidence for scalar runnings only limits the parameter space allowed for higher-order slow-roll parameters to $\xi^2_V=-0.0005\pm0.0050$ and $\varpi^3_V=0.0058^{+0.0063}_{-0.0061}$ both consistent with zero within one standard deviation. Similarly, $\epsilon_3=-0.02\pm 0.26$ while $\epsilon_4$ turns out to be unbounded for all the datasets considered in this work, so we do not show it in the Tables. Adding also lensing or BAO data to Planck, the constraints on the inflationary parameters do not change significantly, see also Table~\ref{tab.LCDM+runnings}.
Interestingly, considering the Atacama Cosmology Telescope DR4 likelihood combined with WMAP 9-years, the bounds on $\eta_V=-0.0160\pm0.0041$ and $\epsilon_2=0.0320\pm0.0082$ remain basically unchanged with respect to the other datasets, while the preference for nonvanishing runnings is translated into the constraints on higher-order inflationary parameters $\xi_V^2=-0.0174\pm0.0058$ and $\varpi^3_V= 0.0172\pm0.0064$ (or equivalently $\epsilon_3<-0.02$ at 95\% CL) that are all different from zero at more than 95\% CL. Finally, regarding the SPT3G+WMAP case, we find more than $1\sigma$ shift toward lower values of both $\eta_V=-0.0111\pm0.0053$ and $\epsilon_2=0.022\pm0.011$, while we find $1\sigma$ preference for nonvanishing higher-order parameters $\xi_V^2=-0.0141\pm0.0085$ and $\varpi^3_V= 0.0115\pm0.0078$. For this dataset $\epsilon_3$ is instead unconstrained.
 

\subsection{Slow-roll relations and the tensor spectrum}

We now analyze a scenario which includes as additional parameters both the running of the scalar tilt $\alpha_s$ and the tensor amplitude, parametrized through the so-called tensor-to-scalar ratio $r$. We refer to this model as $\Lambda\rm{CDM}+\alpha_s+r$. 

In Table~\ref{tab.LCDM+r+nrun} we summarize the results obtained for this model while in Fig.~\ref{fig:LCDM+r+nrun} we show the $68\%$ and $95\%$~CL contour plots for different inflationary parameters.

For the scalar parameters, we see that the constraints on $n_{s}$ and $\alpha_s$ are slightly changed when replacing the running of running with the tensor-to-scalar ratio. This is due to the fact that, once the tensor amplitude varies, also the terms $\propto \epsilon_V$ contribute in the slow roll relations \eqref{Spectral} and \eqref{alpha_s}, modifying the correlation among the inflationary (scalar and tensor) parameters discussed in the previous subsection. Moreover since $\alpha_{s}$ and $\beta_s$ are strongly correlated for all the datasets (see also Fig~\ref{fig:LCDM+runnings}) fixing $\beta_s=0$ produces a shift of $\alpha_s$ toward lower values. In particular, one can see that for the Planck data this shift is translated into a preference for negative values of $\alpha_s$ at the level of slightly more than $1\sigma$ even though the constraints on the running are always consistent with zero within two standard deviations. Notice also that these results remain unchanged when the lensing and BAO measurements are considered together with Planck. Furthermore, when the tensor amplitude can freely vary and the running of running is fixed to zero, also the ACTPol+WMAP and SPT3G+WMAP constraints on $\alpha_s$ shift toward lower values. This produces a reduction of $\alpha_s=0.0090\pm0.0087$ for ACTPol+WMAP, positive and larger than zero at slightly more than one standard deviation, and $\alpha_s=0.001\pm0.012$ for SPT3G+WMAP, completely in agreement with a vanishing scalar running. It should be noticed here that while SPT3G+WMAP is in agreement with Planck for the value of the running $\alpha_s$, ACTPol+WMAP is instead in tension at about $2\sigma$. As in the previous case the difference we see also in Fig.~\ref{fig:LCDM+r+nrun} is coming from the high multipole region.

As concerns the tensor spectrum, we see that its amplitude is constrained to be $r<0.165$ (at 95\% CL) by the Planck data alone while ACTPol+WMAP and SPT3G+WMAP give $r<0.176$ and $r<0.260$, respectively. A strong improvement in this upper bound is obtained including also the BK15 data that, combined with Planck, gives $r<0.0658$. Using the slow-roll relation between the tensor amplitude and the tensor tilt, $n_{\rm T}=-r/8$, these upper bounds on the amplitude can be translated into a lower bounds on the (negative) tensor tilt, namely: $n_{\rm T}>-0.0206$ for the Planck data and $n_{\rm T}>-0.0082$ for Planck+BK15. Furthermore, in the slow-roll framework, any constraint to the tensor amplitude places also a constraint to the energy scale of inflation which reads
\begin{equation}
V^{1/4}_{\rm inf}=M_{\rm pl}\left(\frac{3}{2}\,\pi^2\,A_s\,r\right)^{1/4} \,\rm{GeV}. 
\label{V1/4}
\end{equation}
Using the results in Tab.~\ref{tab.LCDM+r+nrun} from Planck data we derive $V^{1/4}_{\rm inf}<2.04\times 10^{16}\,\rm{GeV}$ while the inclusion of the BK15 data improves this upper bound to $V^{1/4}_{\rm inf}<1.62\times 10^{16}\,\rm{GeV}$.

Reversing the slow-roll relations for the scalar and tensor parameters, we derive constraints on the slow-roll parameters $\{\epsilon_V\,,\eta_V\,,\xi^2_V \}$ that are related to the shape of the inflationary potential. In particular from Planck, we get $\epsilon_V<0.0103$ while the improvement in the constraining power on the tensor amplitude due to the BK15 data is translated into the improved upper bound $\epsilon_V<0.0041$. On the other hand, for $\eta_V$ and $\xi_V^2$ the Planck + BK15 data give $\eta_V=-0.0130^{+0.0038}_{-0.0050}$ and $\xi_V^2=0.0038\pm0.0034$, respectively, ruling out the null value at more than one standard deviation. On the contrary, ACTPol+WMAP finds $\eta_V=0.0015^{+0.0074}_{-0.013}$ and $\xi_V^2=-0.0045\pm0.0044$, always showing $1\sigma$ indication different from zero, but with an opposite sign with respect to Planck. In addition, SPT3G+WMAP prefer both the parameters $\eta_V$ and $\xi_V^2$ in agreement with the null value within the 68\% CL. Equivalently, we can derive constraints on the parameters $\{\epsilon_2\,,\epsilon_3\}$. For Planck + BK15 we obtain $\epsilon_2 = 0.0334\pm 0.0054$ and $\epsilon_3 = 0.24\pm 0.21$. Instead, the Atacama Cosmology Telescope and the South Pole Telescope data, even if they have larger experimental errors and lead to less constraining bounds, prefer $\epsilon_2$ much lower than Planck, reducing the significance for a value different from zero, and $\epsilon_3$ unconstrained.

Under the assumption of slow roll inflation, we see that the parameter space allowed for the (higher-order) tensor parameters in the slow-roll paradigm is strongly reduced since constraints on $r$ and the scalar spectrum are translated into constraints on tensor spectrum, see also Fig.~\ref{fig:LCDM+r+nrun}. In particular using the Planck+BK15 data we see that the results for the scalar parameters and the upper bound on the tensor amplitude, are translated into the constraints $\alpha_T=\left(\,-11.7^{+7.9}_{-5.9}\,\right)\cdot 10^{-5}$ and $\beta_T=\left(\,3.9^{+2.5}_{-4.8}\,\right)\cdot 10^{-5}$ for the tensor running and its running of running. It should be noted that these results are consistent with zero within less than two standard deviations and that, in any case, they are expected to be extremely small and therefore negligible in the slow-roll hierarchy. 
Similar results can be obtained also exploiting the Planck-independent measurements by ACTPol+WMAP and SPT3G+WMAP, see Tab.~\ref{tab.LCDM+r+nrun}. In particular, for these datasets the bounds on $\alpha_{\rm T}$ and $\beta_{\rm T}$ turn out to be less constraining with respect Planck(+BK15) because ACTPol and SPT3G in combination with WMAP have a smaller sensitivity both on the tensor amplitude and on scalar modes. However the higher-order corrections to the power-law spectrum of gravitational waves are always constrained to be extremely small by the slow-roll relations and, given also the large error bars of all the dataset, the bounds are all consistent with each other within 2 standard deviations. This leads to predict a scale invariant tensor tilt, unless corrections of order $|dn_{\rm T}/d\log k|\lesssim 10^{-5}$.


\subsection{Implications for slow roll inflationary models}
Now, we shall focus on the constraints that can be derived on the slow-roll models of inflation. In particular, for a few selected models, we compute the slow-roll parameters and consequently we predict the values of $n_s$, $\alpha_s$ and $r$ to first order in the slow-roll approximation. We include an uncertainty in the number of \textit{e}-folds (before the end of inflation) of $50<N<60$ \cite{Akrami:2018odb}.  In Fig.~\ref{fig:Models} we compare the theoretical predictions with the observational constraints obtained within the $\Lambda\rm{CDM}+r+\alpha_s$ cosmological model for the different datasets listed in Sec~\ref{sec.Methods}.

First, by noting that in a Universe dominated by the energy-density of the inflaton field during the slow-roll regime we have $\dot{H}=-4\pi G\dot{\phi}^2=d^2N/dt^2$, one can relate the field excursion to the tensor amplitude by $\Delta\phi / M_{\rm pl}=\sqrt{r/8}\,N$ and using $N=50$ we set a lower bound
\begin{equation}
   \frac{\Delta\phi}{M_{\rm pl}}=1.01\left(\frac{r}{3.26\times10^{-3}}\right)^{\frac12}
\end{equation}
that is shown in Fig.~\ref{fig:Models}. Note that both large and small field models are compatible with every dataset. Then, using Eq.~\eqref{V1/4}, we get an approximate limit for potentials that work on GUT scales finding that they are ruled out at 95\% CL by the combination Planck+BK15 even though they are still compatible with the other datasets, including ACTPol+WMAP and SPT3G+WMAP. 
This is again an indication of a tension between the Planck satellite results and the ground based telescopes measurements, that prefer a larger value for the scalar spectral index $n_s$ more consistent with a scale invariant spectrum $n_s=1$. This is not only a volume effect, due to the different constraining power of the experiments, but also an actual shift of the $n_s$ constraints coming from the power spectra damping tails.
Lastly, we determine whether the data are in agreement with a convex or a concave potential, being $r=-8/3 \,(n_s-1)$ the relation which defines the limit between the two different shapes. Due to the fact that B-modes polarization measurements are able to give more stringent constraints on tensor modes, in particular on $r$ that appears in the relation aforementioned, the BK15 data indicates that the potential should present a concave shape and exclude completely a convex one, whereas the other datasets are unable to give such a restriction and allow both shapes, see also Fig.~\ref{fig:Models}.

Finally, we present below a concise review of the inflationary models studied in this work and the main results obtained by our analysis. 

\begin{itemize}[leftmargin=*]
	
\item (Generalized) natural inflation: we  start from the general natural inflation \cite{Munoz:2014eqa}, which consider an axion model where a global $U(1)$ symmetry is spontaneously broken at scale $f$, with soft explicit symmetry breaking at a lower scale $\Lambda$; the inflaton field is the pseudo-Nambu-Goldstone boson \cite{Adams:1992bn}. The potential reads
\begin{equation}
    V=2^{1-m}\Lambda^4\left[1+\cos{\frac{\phi}{f}}\right]^m.
    \label{NI}
\end{equation}
Fixing $m=1$ and recovering the natural inflation \cite{Freese:1990rb}, the parameters are
\begin{gather}
    n_s=1-\frac{1}{y}\left[\frac{1+2y^2(1+e^{-x})}{1+2y^2(1-e^{-x})}\right],\\
    \alpha_s=-\frac{4(2y^2+1)e^x}{y^2(-2y^2+(2y^2+1)e^x)^2},\\
    r=\frac{16e^{-x}}{1+2y^2(1-e^{-x})},
\end{gather}
where $x=N/y$ and $y=f/M_{\rm pl}$. Plotting the above quantities as functions of $f/M_{\rm pl}$ (blue curves in Fig.~\ref{fig:Models}) we can see that this model in only compatible within one standard deviation for Planck and within two standard deviation with Planck+BK15. Anyway, relaxing the assumption $m=1$ and leaving $m$ a free parameter, the compatibility with Planck+BK15 increases as long as $m<1$. Given the tension present in the parameter space between the different experiments (as we can see from Fig.~\ref{fig:LCDM+r+nrun}), the model compatibility changes between the datasets. 
In fact, the South Pole Telescope data show only an agreement at 95\% CL for every N in the chosen interval, \ie both blue lines are in the lighter region of the dataset. Moreover, the shift toward high values of $n_s$ preferred by the Atacama Cosmology Telescope data basically excludes the (generalized) natural inflation from the 95\% CL contours. It should be stressed that the ground experiments (ACTPol and SPT-3G) are the ones responsible for the shift of the measurements and consequently changes the compatibility with the model, not WMAP 9-years \cite{2013ApJS..208...19H,Planck:2018nkj}.
Actually, the shift of the $n_s$ bounds is due to the high multipole region accurately constrained by the damping tail of the power spectra.
 
\item (Non minimally coupled) power-law inflation: by taking the limit $f\rightarrow \infty$ in Eq.\eqref{NI}, we recover the quadratic potential, a particular case of the general power-law inflation, represented as two yellow straight lines in Fig.~\ref{fig:Models}, and described by the dominant term $\lambda_n\phi^n$. Values of the index $n=2/3,1,2$ have been obtained in string theory \cite{Silverstein:2008sg,McAllister:2008hb,Dimopoulos:2005ac}. The spectral index, the scalar running and $r$ are simply
\begin{gather}
    n_s=1-\frac{2n+4}{n+4N},\\
    \alpha_s=-\frac{8(n+2)}{(n+4N)^2},\\
    r=\frac{16n}{n+4N}
\end{gather}
and we can see that there is no agreement when the B-modes BK15 observation are included, whereas we still have a consistency at 95\% CL with Planck alone, or up to within $1\sigma$ for ACTPol+WMAP and SPT3G+WMAP. Nevertheless, provided a nonminimal coupling with gravity, the simple power-law potential acquires a compatibility up to 68\% CL for some values of $n$ as shown by the red lines in Fig.~\ref{fig:Models}. The coupling constant $\xi$ is chosen according to Ref.~\cite{Shokri:2019rfi} where the authors have made an analysis imposing this inflationary model at the beginning and using $\xi$ as a free parameter. For the sake of completeness the values are listed below
\begin{itemize}
    \item $\mathbf{n=4}$, with $\xi\simeq 0.0016$,
    \begin{gather}
        n_s=1-\frac{1}{N}(3-8\xi N),\\
        \alpha_s=\frac{1}{N^2}(-3+96\xi N-64\xi^2N^2),\\
        r=\frac{16}{N}(1-8\xi N).
    \end{gather}
    \item $\mathbf{n=2}$, with $\xi\simeq0.0015$,
    \begin{gather}
        n_s=1-\frac{2}{N}(1+\frac43\xi^2N^2),\\
        \alpha_s=\frac{2}{N^2}(-1+4\xi \alpha_sN-96\xi^2N^2),\\
        r=\frac{8}{N}(1-8\xi N).
    \end{gather}
    \item $\mathbf{n=\frac43}$, with $\xi\simeq 0.0011$,
    \begin{gather}
        n_s=1-\frac{1}{3N}(5+8\xi N),\\
        \alpha_s=\frac{5}{81N^2}(-27+48\xi N-704\xi^2 N^2),\\
        r=\frac{16}{9N}(3-32\xi N).
    \end{gather}
    \item $\mathbf{n=2/3}$, with $\xi\simeq 0.0007$,
    \begin{gather}
        n_s=1-\frac{4}{3N}(1+4\xi N),\\
        \alpha_s=\frac{4}{81N^2}(-27+84\xi N+464\xi^2 N^2),\\
        r=\frac{8}{9N}(3-40\xi N).
    \end{gather}
\end{itemize}
This model is consistent also with the ATCPol+WMAP and SPT3G+WMAP contours with the preference for higher values of the tensor tilt translated into slightly preferences for lower values of $n<2$, \textit{e.g.}, the one with $n=2/3$ acquires a compatibility of 68\% CL.

\item Quintessential inflation: in this scenario the early inflationary period and the late-time acceleration are combined. The potential in this case should be shallow at early times, i.e. satisfying the slow-roll conditions, and steep after. As the usual exponential model does not satisfy the observational constraints \cite{Geng:2017mic} a new parameter $n$ is added (\ref{Quint}) which also influences the steepness of $V(\phi)$, whose form is
\begin{equation}
    V=\Lambda e^{-\lambda \frac{\phi^n}{M^n_{pl}}},
    \label{Quint}
\end{equation}
with $n>1$. Imposing $\lambda\ll 1$ we end up with the large field inflation, called quintessential inflation \cite{Peebles:1998qn}. In this model, the parameters are
\begin{gather}
    n_s=1-\frac{2(n-1)}{(n-2)N}-\frac{[n(n-2)\lambda N]^{-\frac{2}{n-1}}}{(n-2)^2N^2},\label{nsExpo}\\
    \alpha_s=-\frac{2(n-1)}{(n-2)N^2}+\frac{6(n-1)[n(n-2)\lambda N]^{-\frac{2}{n-2}}}{(n-2)^3 N^3},\\
    r=\frac{8[n(n-2)\lambda N]^{-\frac{2}{n-2}}}{(n-2)^2N^2}.
\end{gather}
Fixing $\lambda=10^{-10}$ and varying $n$, the purple curves in Fig.~\ref{fig:Models} are drawn, showing, for example as reference, that $n=7$ is compatible with 95\% CL of Planck+BK15 with $N=60$ whereas it is not for $N=50$. A lower values of $\lambda$ move the curves to the right, increasing the inclination, whereas a higher value makes $n_s$ independent of it, as shown in (\ref{nsExpo}). Concerning the other datasets, this model is in disagreement with ACTPol+WMAP data unless for significantly lower values of $\lambda$, and in tension with SPT3G+WMAP. Also in this case the different agreement of the models with the data is affected by the inconsistency between the datasets explored here.

\item Starobinsky-like Inflation: lastly, we analyze the $R^2$ inflation \cite{Starobinsky:1980te} which is characterized by adding higher curvature corrections ($R^2$) to the Hilbert action of gravity \eqref{minimal coupled action}. This analysis comprehends also Higgs inflation \cite{Bezrukov:2007ep} and universal attractors models \cite{Kallosh:2013lkr} predictions since they are equal to the Starobinsky inflation \cite{Kehagias:2013mya}. The similarity is due to the fact that kinetic terms are negligible during the inflationary period and the slow-roll parameters differ from one another for $\sim 10^{-5}$ corrections which are still too small to be measured. However deviations from this scenario can be obtained considering different classes of inflationary models like $\alpha$-attractors \cite{Kallosh:2013tua}. The Starobinsky potential is
\begin{equation}
    V=\frac{M_{\rm pl}^2}{8}\lambda(1-e^{-\sqrt{\frac23}\frac{\phi}{M_{\rm pl}}})^2
\end{equation}
and the inflationary parameters are
\begin{gather}
    n_s=1-\frac{32N+24}{(4N-3)^2}\simeq 1-\frac{2}{N},\\
    \alpha_s=-\frac{64N(8N-13)}{(4N-3)^4}\simeq -\frac{2}{N^2},\\
    r=\frac{192}{(4N-3)^2}\simeq\frac{12}{N^2}.
\end{gather}
In this model the hierarchy of the parameters is $\xi\sim\epsilon\ll\eta\ll1$ instead of the more common $\xi\ll\eta\ll\epsilon\ll1$. Thus the value of $r$ is expected to be extremely small. In fact, we can see from the cyan line in Fig.~\ref{fig:Models} that its smallness results in a compatibility within one standard deviation for all the datasets. Small deviation from this model, i.e. considering the term $R^p$ with $p\approx2$ \cite{Motohashi:2014tra}, worsen the agreement with Planck+BK15 as shown by the dotted lines which represent terms with $2+\Delta p$ where $\Delta p=0.01$. Considering also ACTPol+WMAP, we see that the model is excluded when $p$ is decreased, whereas for SPT3G+WMAP it is  still consistent within the 95\% CL contours. On the other hand, a bigger value of $p$ is completely in agreement with both datasets at 68\% CL.
\end{itemize}

We would like to conclude this subsection with some final remarks. First, we can appreciate that the constraints on the slow roll inflationary models remain basically stable when $dn_s /d\log k$ can freely vary in the sampling, see also the analogous discussion in \cite{Akrami:2018odb} and also \cite{Geng:2017mic,Wu:2018vuj,Renzi:2019ewp,Civiletti:2020fkm,Meza:2021xuq}. However, the tension~\citep{Handley:2020hdp} present between the cosmological datasets analyzed in this work (i.e., Planck, ACTPol+WMAP and SPT3G+WMAP) produces different constraints on the inflationary parameter and consequently completely different results regarding the model compatibility, see also Fig.~\ref{fig:Models}.


\subsection{Inflation and spatial curvature} 

In this section we study two different extensions of the standard cosmological model that both include the curvature parameter $\Omega_k$ as an additional parameter. In particular we first analyze the case $\Lambda\rm{CDM}+r+\Omega_k$ and then we add also the running of the scalar tilt, $\Lambda\rm{CDM}+r+\alpha_s+\Omega_k$. For both the models, we adopt the common power-law parameterization for the primordial spectra, assuming the usual slow-roll consistency relations to hold. Indeed, since the vast majority of inflationary models predict flatness, the constraints on the spatial curvature provide an important consistency check of this standard scenario, see also \cite{Akrami:2018odb}.

Tab.~\ref{tab.LCDM+r+omegak} summarizes the constraints derived for the model $\Lambda\rm{CDM}+r+\Omega_k$ and in Fig.\ref{fig:LCDM+r+omegak} we show the $68\%$ and $95\%$~CL marginalized contours for different inflationary parameters in the same model. On the other hand, in Table~\ref{tab.LCDM+r+nrun+omegak} we present the results for the $\Lambda\rm{CDM}+r+\alpha_s+\Omega_k$ model showing in Fig.\ref{fig:LCDM+r+nrun+omegak}  the $68\%$ and $95\%$~CL contours. 

For the inflationary parameters we see that in both the models, slightly higher values for scalar tilt are preferred with respect to the case without $\Omega_k$. In particular the Planck data gives $n_s=0.9720\pm 0.0052$ ($n_s = 0.9728\pm 0.0052$) when the running $\alpha_s$ is included (excluded). We can also appreciate that these constraints have $1\sigma$ shift toward higher values for the different datasets, including ACTPol+WMAP and SPT3G+WMAP. As concerns the scalar running, the bounds on $\alpha_s$ are consistent with those derived without considering $\Omega_k$, see also Table \ref{tab.LCDM+r+nrun+omegak}.

For the tensor amplitude, we see that, ignoring the scalar running, Planck data gives $r<0.170$ at 95\% CL while including $\alpha_s$ this bound is less stringent: $r<0.250$. Interestingly, for ACTPol+WMAP the upper bound $r<0.210$ becomes more stringent ($r<0.185$) including $\alpha_s$. We also confirm that the ACTPol+WMAP preference for a non-vanishing scalar running is reduced when the tensor amplitude can freely vary. A strong improvement in the constraining power is clearly obtained including also the B-modes BK15 likelihood and, in fact, including (excluding) the running, the combination Planck+BK15 gives $r<0.0637$ ($r<0.0613$). Also in this case the results appear to be stable and consistent with the case in which $\Omega_k$ is not varied.

Using the slow-roll consistency relations among the inflationary parameters, we can appreciate how also in this case the parameters space allowed for the tensor spectrum is strongly constrained. On the other hand reversing the slow-roll relations for the scalar and tensor parameters, we can derive constraints on the slow-roll parameters $\{\epsilon_V\,,\eta_V\,,\xi^2_V \}$. 
Exploiting the Planck+BK15 data, for the $\Lambda\rm{CDM}+r+\Omega_k$ model we obtain $\epsilon_V<0.0038$ and $\eta_V=-0.0094^{+0.0038}_{-0.0049}$ such results remain similar even if we let the scalar running $\alpha_s$ free to vary, in this scenario, however, we have also the result for the slow-roll parameter of the third order: $\xi_V^2=0.0013\pm0.0034$. Considering the ACTPol+WMAP and SPT3G+WMAP datasets combination, we find instead both $\epsilon_V$ and $\eta_V$ in agreement with zero within the 68\% CL when the scalar running is fixed to zero or free to vary, while it appears $1\sigma$ indication for a negative $\xi_V^2$ for ACTPol+WMAP in the $\Lambda\rm{CDM}+r+\alpha_s+\Omega_k$ model.
Equivalently, we can constrain the parameters $\{\epsilon_i\}$ obtaining $\epsilon_2=0.0256\pm 0.0057$ ($\epsilon_2=0.0252\pm 0.0055$) and $\epsilon_3=0.10\pm 0.29$ when $\alpha_s$ is considered (excluded) for Planck+BK15. This indication for the $\epsilon_2$ parameter different from zero is reduced to more than $1\sigma$ for ACTPol+WMAP and disappears for SPT3G+WMAP. We would like to stress that all the results obtained analyzing the Planck 2018 data are in agreement with the ACTPol+WMAP and SPT3G+WMAP data within the 95\% CL.

Interestingly, as concerns the spatial curvature, the Planck preference for a closed universe~\cite{Aghanim:2018eyx,DiValentino:2019qzk,Handley:2019tkm,DiValentino:2020hov} is confirmed in both the scenarios, and slightly enforced when the BK15 data are combined together with Planck Data. Indeed in the extended parameter space of $\Lambda\rm{CDM}+r+\Omega_k$ we obtain $\Omega_k=-0.048^{+0.020}_{-0.016}$ for Planck and $\Omega_k=-0.047^{+0.018}_{-0.015}$ for Planck+BK15. Considering also the running of the scalar tilt as an additional parameter, the results are essentially unchanged. In any case, Planck and Planck+BK15 data prefer $\Omega_k<0$ at $2.4\,\sigma$ and $2.6\,\sigma$, respectively. Anyway, considering the lensing spectrum as measured by the Planck Collaboration the evidence for $\Omega_k\ne0$ is reduced to less then two standard deviation ($\Omega_k=-0.0123^{+0.0072}_{-0.0063}$ and $\Omega_k=-0.0113\pm 0.0066$ ignoring and considering $\alpha_s$, respectively). Finally, we have the indication for a spatially flat universe using also the BAO data ($\Omega_k=0.0007\pm0.0020$, for both the models), but this result should be considered with caution because these measurements are in strong disagreement with Planck when the curvature parameter is free to vary~\cite{DiValentino:2019qzk,Handley:2019tkm,DiValentino:2020hov}, so they cannot in principle be combined together. Similarly, exploiting the data from the Atacama Cosmology Telescope and the South Pole Telescope we do not find any evidence for $\Omega_k\ne 0$, with the constraints reading $\Omega_k=-0.007^{+0.016}_{-0.012}$ ($\Omega_k=-0.010^{+0.017}_{-0.011}$) for ACTPol+WMAP and $\Omega_k=-0.0008^{+0.013}_{-0.0097}$ ($\Omega_k=-0.000^{+0.015}_{-0.011}$) for SPT3G+WMAP when the running is excluded (included).
It is important to stress here, that also in these extended scenario including a curvature free to vary the ACTPol+WMAP and SPT3G+WMAP dataset combinations show a tension with respect to the results obtained by Planck, as we can see in Figs.~\ref{fig:LCDM+r+omegak} and~\ref{fig:LCDM+r+nrun+omegak}, always driven by the same effect discussed before.

So, albeit the Universe is spatially flat or closed is still a very disputed issue, see also \cite{DiValentino:2020srs,Vagnozzi:2020zrh,Vagnozzi:2020dfn,Dhawan:2021mel,Efstathiou:2020wem,Yang:2020bpv}, in what follows we take into account the Planck(+BK15) preference for a closed cosmological spacetime, investigating the possible consequences for the slow-roll background dynamics. 

Inflation in a curved Universe has been largely discussed in literature, see e.g. Refs~\cite{Linde:1995xm,Linde:2003hc,Ratra:2017ezv,Bonga:2016iuf,Handley:2019anl,Bonga:2016cje,Ooba:2017ukj,Ellis:2001ym,Uzan:2003nk,Unger:2018oqo,Gordon:2020gel,Sloan:2019jyl,Motaharfar:2021gwi}. As a matter of fact, during inflation the spatial curvature is exponentially driven to flatness and so the only way to obtain an inflationary universe with $\Omega_k\ne0$ is to assume that it inflated only by a finite (small) number of \textit{e}-folds $\Delta N_{\rm tot}$. Furthermore, in a curved inflationary background, the power-law relations adopted in this work to compute the primordial spectra become disputed at low multipoles $\ell\lesssim 20$ and more reliable parameterizations should be considered~\cite{Ratra:2017ezv,Bonga:2016iuf,Handley:2019anl,Bonga:2016cje}. Anyway the differences are typically limited to low multipoles and the Planck estimation of cosmological parameters remains robust under the inclusion of positive spatial curvature~\cite{Bonga:2016iuf}. In what follows we therefore neglect these corrections and we provide constraints on the \textit{e}-fold of inflation compatible with Planck(+BK15) preference for a closed Universe. Indeed, in the case of a positive curvature, $\Omega_k<0$, assuming a slow-roll evolution and a reheating phase taking place just after the end of inflation ($\rho_{\rm reh}\simeq V_{\rm inf}$), the total of \textit{e}-fold can be estimated as~\cite{Ellis:2001ym,Uzan:2003nk} 
\begin{equation}
\Delta N_{\rm tot}\simeq \frac{1}{2}\log\left(\frac{\left(1+\delta_0-\Omega_{\rm rad}\right)\mathcal R + \Omega_{\rm rad} \mathcal R^2}{\delta_0}\right)
\end{equation}
with $\delta_0=\Omega_0-1$, $\Omega_{\rm rad}\simeq 4\times 10^{-5}\,h^{-2}$ the radiation density parameter today~\cite{Aghanim:2018eyx} and 
\begin{equation}
\log \mathcal R \simeq 66 + \log\left(\frac{V^{1/4}_{\rm inf}}{10^{16}\,\rm{GeV} }\right).
\end{equation}
In Fig.~\ref{fig:Ntot_exit}, we show the 68\% and 95\% CL marginalized contours for the total number of \textit{e}-fold of inflation compatible with Planck(+BK15) preference for a closed Universe. Within the $\Lambda\rm{CDM}+r+\Omega_k$ model, using only the Planck data, we obtain a maximum number of \textit{e}-fold $\Delta N_{\rm tot}=63.55^{+0.30}_{-0.21}$ at 68\% CL while including also the B-modes likelihood, for Planck+BK15 we get $\Delta N_{\rm tot}=63.31^{+0.31}_{-0.23}$ at 68\% CL. Including the scalar running in the sampling, the results remain almost unchanged, see also Tab~\ref{tab.LCDM+r+nrun+omegak} and Fig.~\ref{fig:Ntot_exit}. This means that if the Planck(+BK15) evidence for a closed Universe will be confirmed by future measurements, one would need about 63 \textit{e}-fold of expansion while the total number of \textit{e}-folds in many physical models of inflation is typically extremely large, e.g. in power-law inflation one expects $\Delta N_{\rm tot}\sim 10^{12}$ \cite{Lyth:1998xn,Linde:2005ht}. This would strongly constrains the background dynamics before the largest observable scale exit the horizon, with important implications for the observed homogeneity in the cosmic microwave background. Indeed, assuming a standard slow roll inflation followed by a canonical reheating phase and supposing the Universe to be radiation-dominated from the end of reheating to the matter-radiation equality, the number of \textit{e}-folds between when the scale $k$ crosses the horizon and the end of inflation can be estimated as~\cite{Uzan:2003nk,Liddle:1993fq,Lidsey:1995np}
\begin{align}
\nonumber N(k)\simeq & 128 - \log \mathcal R-\log \left(\frac{k}{a_0\,H_0}\right) +2\log\left(\frac{V^{1/4}_{\rm inf}}{10^{16}\,\rm{GeV} }\right)\\& - \log\left(\frac{H_0}{100\,\rm{Km/s/Mpc}}\right) + \mathcal O \left(\log (V_k / V_{\rm inf})\right)
\end{align}
where, for a slow-roll dynamics, the effects of assuming  $V_k \simeq V_{\rm inf}$ are expected to be small for the scales of interest. By noting that the CMB roughly probes scales from 10 to $10^4$ Mpc, one can estimate the number of \textit{e}-fold before the largest observable scale in the Universe exits the horizon $\Delta N(k_{\rm exit})\simeq \Delta N_{\rm tot} - N(k_{\rm min})$. By noting that for the parameter space explored in this work $N(k_{\rm min})\simeq 61 - 62$, see also \cite{Uzan:2003nk}, from Planck(+BK15) data it follows that, within the $\Lambda\rm{CDM}+r+\Omega_k$ model, $\Delta N(k_{\rm exit})=1.55^{+0.30}_{-0.21}$ ($\Delta N(k_{\rm exit})=1.31^{+0.31}_{-0.23}$), while including also $\alpha_s$ we get $\Delta N(k_{\rm exit})=1.67^{+0.29}_{-0.21}$ ($\Delta N(k_{\rm exit})=1.34^{+0.30}_{-0.22}$), see also Fig.~\ref{fig:Ntot_exit}. Although the allowed number of \textit{e}-fold compatible with the constraints by structure formation (i.e., 50 - 60 \textit{e}-folds between the horizon exit and the end of inflation \cite{Akrami:2018odb}) are enough also to solve ‘flatness’ with an accuracy represented by the precision in $\Omega_k$ (a fine tuning of about 1\% is typically enough \cite{Linde:2003hc}), it should be also noted that the main difficulty for a successfully closed inflationary model is represented by homogeneity and isotropy. Indeed, in most of the models proposed in the literature, when the universe does not inflate long enough to become flat, the density perturbations on the horizon scale are typically expected to be much larger than those observed, except for a specific class of models~\cite{Linde:2003hc}.

\section{Conclusion}\label{sec.Conclusion}

In this paper, exploiting the most recent cosmological observations, we provide new updated constraints on slow roll inflation analyzing different extended scenarios beyond the $\Lambda\rm{CDM}$ cosmological model. Together with the usual six standard parameters, we simultaneously vary different combinations of additional parameters, including the running of the scalar spectral index $\alpha_s\doteq dn_s/d\log k$, its running of running $\beta_s\doteq d\alpha_s / d\log k$, the tensor amplitude $r\doteq A_T / A_s$ and the spatial curvature $\Omega_k$. Our baseline dataset consists of Planck 2018 data and its combination with the Planck 2018 lensing likelihood , the baryon acoustic oscillations (BAO) measurements and the B-modes BK15 likelihood. We also analyze two other independent cosmological datasets provided by the Atacama Cosmology Telescope (ACTPol) and South Pole Telescope (SPT3G) both combined with WMAP 9-years observations data.

As concerns the spectrum of primordial scalar perturbations, analyzing the different combinations of the Planck, lensing, BAO and BK15 data, we find no evidence for a scalar running or a running of running. On the other hand, analyzing the ACTPol+WMAP data we find a preference for nonzero $\alpha_s$ and $\beta_s$ at the level of $2.9\sigma$ and $2.7\sigma$, respectively (\textit{i.e.}, both at about 99\% CL). Anyway, such a preference is reduced when the running of running is replaced by tensor amplitude in the model.  

Regarding the spectrum of inflationary gravitational waves, we provide different upper bounds on the tensor amplitude, with $r<0.0658$ at 95\% CL our most constraining bound for Planck+BK15 data at the pivot scale $k_{\star}=0.05\rm{Mpc}^{-1}$. This result remains stable in almost all the models considered in the work, implying for the inflationary energy scale $V^{1/4}_{\rm inf}\lesssim 2 \times 10^{16}\,\rm{GeV}$. Furthermore, given the constraints on the tensor spectrum and the upper limits on the tensor amplitude, we show that the slow-roll consistency relations strongly reduce the parameter space allowed for the tensor spectrum, basically predicting a scale-independent tensor tilt unless corrections of order $d n_{T}/d\log k \lesssim 10^{-5}$, for all the datasets.
 
Always using the slow-roll relation, we provide constraints on the slow-roll parameters $\{\epsilon_V \, , \, \eta_V \, , \, \xi_V^2 \, , \, \varpi^3_V\}$ that define the shape of the inflationary potential or similarly on the parameters $\{\epsilon_i\}$ that define the dynamics of the background expansion. We then compare the theoretical predictions of some selected inflationary models with the Planck (+ lensing , BAO, and BK15) ACTPol+WMAP and SPT3G+WMAP data leaving $d n_s /d\log k \ne 0$ free to vary. While the inclusion of the scalar running does not drastically change the constraints on the inflationary models, it should be instead noted that the shift toward higher values of $n_s$ preferred by the ACTPol+WMAP data actually strongly disfavors some inflationary models that are instead compatible with Planck bounds; in some cases leading to completely different conclusions for the model selection.
In other words, the extensions of the standard model considered in this paper recast the global tension between the datasets already present for a $\Lambda$CDM model~\citep{Handley:2020hdp} analysis on a difference between the inflationary parameters.

Finally, we vary the spatial curvature parameter $\Omega_k$ in realistic inflationary models that include both a tensor amplitude and a scalar running. Since the vast majority of inflationary models predict flatness, the constraints on the spatial curvature provide an important consistency check of this standard scenario. Interestingly, analyzing Planck(+BK15) data we instead find a preference for a closed cosmological spacetime at $2.4\,\sigma$ ($2.6\,\sigma$), while no relevant evidence is obtained adding also lensing and BAO to Planck or analyzing the Atacama Cosmology Telescope and the South Pole Telescope data. Maintaining an agnostic perspective on the spatial geometry, we investigate the possible consequences of a curved cosmological spacetime for the inflationary slow-roll background dynamics. Without assuming any underlying physical realization of closed inflation, under the assumption of a canonical slow-roll evolution, we derive constraints on the total number of \textit{e}-fold compatible with the Planck(+BK15) indication for a closed cosmological spacetime, obtaining $\Delta N_{\rm tot}= 63.67^{+0.29}_{-0.21}$ ($\Delta N_{\rm tot}=63.33^{+0.30}_{-0.22}$) at 68\% CL. We also argue that this would strongly limit the allowed \textit{e}-fold of inflation before the largest observable scale exits the horizon to $\Delta N(k_{\rm exit})\lesssim 2$ and we briefly point out the implications for homogeneity.

\section*{Acknowledgements}

M.F., W.G. and A.M. are supported by "Theoretical Astroparticle Physics" (TAsP), iniziativa specifica INFN. E.D.V. acknowledges the support of the Addison-Wheeler Fellowship awarded by the Institute of Advanced Study at Durham University. We thank Andrei Linde and Renata Kallosh for the useful comments and suggestions. 
In this work we made use of the following \texttt{python} packages that are not mentioned in the text : \texttt{SciPy}~\cite{Virtanen:2019joe} for numerical sampling of the statistical distributions involved in our data analysis, \texttt{GetDist}~\cite{Lewis:2019xzd} a tool for the analysis of MCMC samples which employs \texttt{Matplotlib}~\cite{Matplotlib} for the realization of the plots in the paper and \texttt{NumPy}~\cite{NumPy} for numerical linear algebra. All the data underlying this article are publicly available.

\bibliography{main}
\end{document}